\documentclass[a4paper]{aa}
\usepackage{amssymb, longtable}
\usepackage{graphicx}
\usepackage{babel}
\usepackage{hyperref}
\usepackage{soul}
\usepackage{color}
\usepackage{enumerate}
\usepackage{natbib}
\bibpunct{(}{)}{;}{a}{}{,} 
\usepackage{floatrow}
\newfloatcommand{capbtabbox}{table}[][\FBwidth]

\begin{document}

\newcommand{\CaAX}{$A^2\Pi_r - X^2\Sigma^{+}$} 
\newcommand{\CaBX}{$B/B^\prime 2\Sigma^{+} - X^2\Sigma^{+}$} 
\def\AX{$A~^2\Pi~-~X~^2\Sigma^+$~}
\def\pr{Proxima}
\def\ga{G~224-58~A~}
\def\Tef{$T_{\rm eff}$~}
\def\Teff{$T_{\rm eff}$}
\def\Vr{$V_{\rm r}$~}
\def\Vsini{$v$ sin $i$~}
\def\Vm{$V_{\rm m}$~}
\def\EE{$E^{\prime\prime}$~}
\def\kmps{kms$^{\rm -1}$~}
\def\um{$\mu$m~}
\def\logg{$\log~g$~}
\def\loggg{$\log~g$}
\def\Ha{$H_{\alpha}~$}
\def\pew{$pEW_b~$}
\def\vlt{VLT/X-shooter~}
\def\Po{P$_o$}

\title{Flare activity and photospheric analysis of Proxima Centauri}

\author{
Y.\ Pavlenko \inst{1,2,3}
A. Su\'arez Mascare\~no \inst{4,5,7}, 
R.\ Rebolo \inst{4,5,6},
N.\ Lodieu \inst{4,5},
V. J. S. B\'ejar \inst{4,5},
J.I. Gonz\'alez Hern\'andez\inst{4,5} 
}
\institute{Main Astronomical Observatory of the National Academy of Sciences of Ukraine.
         \email{yp@mao.kiev.ua}
         \and
         Instituto de Astrof\'isica de Canarias (IAC), La Laguna, Tenerife, Spain
         \and
          Center for Astrophysics Research, University of Hertfordshire, College Lane, Hatfield, Hertfordshire AL10 9AB, UK
         \and
Instituto de Astrof\'isica de Canarias (IAC), Calle V\'ia L\'actea s/n, E-38200 La Laguna, Tenerife, Spain. 
         \and
         Departamento de Astrof\'isica, Universidad de La Laguna (ULL), E-38205 La Laguna, Tenerife, Spain.
         \and
          Consejo Superior de Investigaciones Cient\'ificas, CSIC, Spain, 
          \and
          Observatoire Astronomique de l'Universit\'e de Gen\`eve, E-1290 Versoix, Gen\`eve, Switzerland.
}

\offprints{Yakiv Pavlenko}
\mail{yp@mao.kiev.ua}

\date{\today{}}

\authorrunning{Pavlenko et al.}
\titlerunning{Spectrum analysis of Proxima Cen}

\abstract{
We present the analysis of emission lines in high-resolution optical spectra 
of the planet-host star Proxima Centauri (Proxima) classified as a M5.5V\@.  
}{
We carry out the detailed analysis of observed
spectra to get a better understanding of the physical conditions of the atmosphere of this star.
}{
We identify the emission lines in a serie series of 147 high-resolution optical
spectra of the star at different levels of activity and compare them with the synthetic spectra
computed over a wide spectral range.
}{
Our synthetic spectra computed with the PHOENIX 2900/5.0/0.0 model atmosphere 
fits pretty well the observed optical-to-near-infrared spectral energy distribution. 
However, modelling strong atomic lines
in the blue spectrum (3900--4200\AA{}) requires implementing additional opacity.
We show that high temperature layers in Proxima Centauri consist in at least 
three emitting parts:  a) a stellar chromosphere where numerous emission 
lines form. We suggest that some emission cores of strong absorption 
lines of metals form there; 
b) flare regions above the chromosphere, where hydrogen Balmer lines up to high transition levels (10--2) form; 
c) a stellar wind component with V${r}$\,=\,$-$30 \kmps{} seen in some Balmer lines
as blue shifted emission lines. We believe that the observed He line at 4026\AA{} 
in emission can be formed in that very hot region.
}{
We show, that real structure of the atmosphere of Proxima is rather complicated. 
The photosphere of the star is best fit by a normal M5 dwarf spectrum. On the other hand
emission lines form in the chromosphere, flare regions and extended hot envelope.
}

\maketitle 

%
%
\section{Introduction}

M dwarfs are the most numerous and longest-lived stars in 
our Milky Way, see \cite{kirkpatrick12}.  Unfortunately, the 
determination of the basic parameters of these stars is hampered by the complicated 
physical processes taking place in their atmospheres which limit our ability to reproduce 
their spectra with synthetic models. 
Due to the low temperatures and high pressures in M dwarf photospheres, modelling their 
spectra requires 
detailed accounting for molecules when dealing with chemical equilibrium in their atmospheres.
 M dwarf spectra are governed by absorptions of the numerous band systems 
of diatomic and poly-atomic molecules. Spectra of M dwarfs also  show emission lines 
which can be formed only in the outermost high temperature layers of their atmospheres.
 
Proxima Centauri (=\,2MASS\,J14294291$-$6240465; GJ\,551, V645 Cen) is the closest red dwarf 
to the Sun located at a distance of 1.3019$\pm$0.0018 pc \citep{lurie14}. Because of
its proximity, its angular radius can be measured directly via interferometry \citep{kervella03}.
Its mass is about an eighth of the Sun's mass, its luminosity is only 0.15\% of that 
emitted by the Sun, its spectral type is M5.5 \citep{bessell91}, its effective temperature 
is 3050\,K, and its density about 40 times that of our Sun. 
Since its discovery, Proxima Centauri has been suggested to be the third component of the
 $\alpha$ Centauri system.
Recently, \cite{kerv17} based on new observations claim that Proxima and $\alpha$ 
Cen are gravitationally bound with a high degree of confidence.
It may be the third component
of the Alpha Centauri system with a projected physical separation of 15,000$\pm$700\,au
\citep{wertheimer06}.

Proxima Cen is a known flare star that exhibits random but significant increases in brightness 
due to magnetic activity \citep{christian04a}. The
spectrum of Proxima Cen contains numerous emission lines, see \cite{fuhr11}.
These features most likely originate from plage, spots, or a combination of both.
In general, the flare rate of Proxima Cen is lower than that of other flare stars 
of similar spectral type, but is unusually high given its slow rotation period \citep{dave16}.
The star has an estimated rotation period of $\sim$ 83 days and a magnetic cycle of 
$\sim$7 years \citep{bene98, suar15, suar16, warg17}.
The X-ray coronal and chromospheric activity have been studied in detail by \citet{fuhr11} 
and \cite{warg17}. Recently \cite{thom17} claimed detection of rotational modulation of emission lines in 
the Proxima Cen spectrum. 

Nowadays M dwarfs represent  important targets for searches of exoplanets, and in particular, 
rocky planets. Given their small radius and low mass, planets are easier to detect around
M dwarfs because the depth of their transits and the amplitudes of the induced
radial velocity variations are larger. First rocky planets were detected by radial velocity 
and transits around M stars \citep{rive15, char09}. Most of the rocky planets in the habitable 
zone have been found around these very low-mass stars 
\citep{udry07,bonf11,quin14,torr15,wrig16,anglada_escude16}.
Proxima Cen was recently highlighted as a planet host mid-M dwarf \citep{anglada_escude16}. 
Proxima Cen b orbits its host star with a period of 11.2 days, corresponding to a semi-major 
axis distance of 0.05 AU\@. Proxima Cen\,b has a mass close to that of the Earth (from 1.10 
to 1.46 mass of the Earth), with orbit in the temperate zone \citep{anglada_escude16}.

One may assume that Proxima Cen b is surrounded by an atmosphere with a surface 
pressure of one bar, implying that the planet orbits its host star within the habitable zone 
 \citep{riba16, turb16, garr16}. It is worth noting that classical definition of habitable zone 
solely implies the restriction of distance from the central star and composition of the 
planetary atmosphere. However, strong flare activity 
may move the inner boundary of the habitability  zone far away from the formally computed 
possible radius. 
For this reason, the detailed characterisation of the flare phenomenon present in the 
atmosphere of Proxima Cen is of great importance.

In this paper we report on the detection of emission lines in the optical spectra of
Proxima taken with different instruments ran by the European Southern Observatory (ESO). 
In Section \ref{ProxCen:spec_obs}, we describe the observations and data reduction. 
In Section \ref{ProxCen:results} we analyse the characteristics of several lines, including 
Balmer lines, sodium resonance  doublet  as well as the Ca II H and K lines.
In Section \ref{ProxCen:discussion}, we place our results into a wider context of activity 
in low-mass stars.

%
%
\section{Spectroscopic observations \label{ProxCen:spec_obs}} 

\subsection{3.6-m/HARPS optical spectra of Proxima}

 We retrieved all the available of Proxima spectra from the HARPS ESO public data archive. 
The dataset consists in 316 spectra collected between June 2004 and May 2016. 
HARPS \citep{Mayor2003} is a fibre-fed high resolution echelle spectrograph installed at 
the 3.6-m ESO telescope in La Silla Observatory (Chile). The instrument has a resolving 
power R$\sim$115\,000 over a spectral range from 3780 to 6810\AA{} and has been 
designed to attain very high long-term radial velocity (RV) accuracy. It is 
contained in a vacuum vessel to avoid spectral drifts due to temperature and air pressure 
variations, thus  ensuring its stability. HARPS is equipped with its own pipeline 
providing extracted and wavelength-calibrated spectra, as well as RV measurements 
and other data products such as cross-correlation functions and their bisector 
profiles. In order to avoid contamination  of the stellar spectra by the 
calibration lamp we relied only on those spectra taken without simultaneous 
calibration. The final selection consisted in 147 high resolution spectra taken from 
the public ESO archive, observed between 2004 and 2016 with exposure times 
ranging from 450 to 1200 s.  

For the analysis we use the reduced wavelength-calibrated spectra produced by
the HARPS pipeline. We correct every spectrum from the velocity of the star 
and created a high signal-to-noise spectrum by co-adding all the available spectra.

\subsection{Spectra of Proxima in different states of activity} \label{harps_s_qc}

We need very high signal-to-noise spectra in order to perform a detailed study 
 of the activity processes and spectral features of \pr. To do so we create two 
 high S/N spectra with two different sets of individual spectra. One by 
 co-adding all the available spectra, once set in the barycentric frame of 
 reference and corrected from the radial velocity of the star, which gives 
 us a final spectrum containing the information of the spectral features both 
 in times of strong and weak activity of Proxima. We label the resulting 
 spectrum as 'S'. Then we create a second spectrum for which we filter 
 out the spectra obtained during flares, and create a mean spectrum representing 
 the times of quietness of the star by once again co-adding the selected spectra. 
 We label this spectrum as 'QC'. {\textbf Flares are identified by measuring 
 unusually high levels of Ca II H\&K emission and \Ha{} emission. We measure 
 the Mount Wilson S index \citep{noye84}
  and the \Ha{} index defined by \cite{gome11}
  following the procedure illustrated in \citet{suar15}. Spectra that show a S 
  index of \Ha{} index exceeding the seasonal mean by more than 3 times the RMS 
  of the whole series are considered in flare state.} As a result of the 
  process we obtain two 
 spectra with SN $> 100$ in a wide spectral range in two different states 
 (average state and low activity state) which allows us to study the changes 
 in its chromosphere related to changes in its activity level.

\subsection{VLT/X-shooter spectra}

X-Shooter is a multi wavelength cross--dispersed echelle spectrograph 
\citep{dOdorico06,vernet11} mounted on the Cassegrain focus of the
Very Large Telescope (VLT) Unit 2\@. The spectrograph is made 
of three arms covering simultaneously the ultraviolet (UVB; 3000--5500\AA{}),
visible (VIS; 5500--10000\AA{}), and near--infrared (NIR; 10000--24800\AA{}) wavelength
ranges thanks to the presence of two  diachronic splitting the light. The spectrograph is
equipped with three detectors: a 4096$\times$2048 E2V CCD44-82, a 4096$\times$2048
MIT/LL CCID\,20, and a 2096$\times$2096 Hawaii 2RG for the UVB, VIS, and NIR arms,
respectively. 

We downloaded public data of Proxima from the European Southern Observatory
(ESO) science archive. The VLT/X-shooter spectra were taken on 15 January 2014 between
UT\,=\,6h55 and UT\,=\,7h as part of ESO  program 092.D-0300\@. The observation
strategy was 2AB cycles of 12s, 33s, and 44s in the UVB, VIS, and NIR arms, respectively.
The slits of 0.5 arcsec 0.4 arcsec, and 0.4 arcsec were used, yielding resolving powers of 
9900 (3.2 pixels per full-width-half-maximum), 18200 (92.9 pixels per 
full-width-half-maximum), and 10500 (2.2 pixels per full-width-half-maximum) 
in the UVB, VIS, and NIR arms, respectively.
The read-out mode was set to 400k and low gain without binning. 

We reduced the raw dataset with the latest version of the X-shooter 
pipeline (2.8.0)\footnote{http://www.eso.org/sci/software/pipelines/}. 
The pipeline removes the instrumental signature to the raw spectra, including bias and flat-field.
The spectra are wavelength-calibrated, sky-subtracted and finally flux-calibrated 
with the
associated spectro-photometric standard star observed as part of the ESO calibration plan.
The output products include a 2D spectrum associated with a 1D spectrum. Nonetheless,
we extracted the 1D UVB, VIS, and NIR spectra with the {\tt{apsum}}
task under IRAF\footnote{IRAF is distributed by the National Optical Astronomy Observatories, 
    which are operated by the Association of Universities for Research 
    in Astronomy, Inc., under cooperative agreement with the National 
    Science Foundation.} \citep{tody86,tody93}.

%
\section{Results}
\label{ProxCen:results}

\subsection{Absorption spectra of Proxima}

\subsubsection{Effective temperature from the X-Shooter's SED}

\begin{figure*}
  \centering
  \includegraphics[width=0.48\linewidth, angle=0]{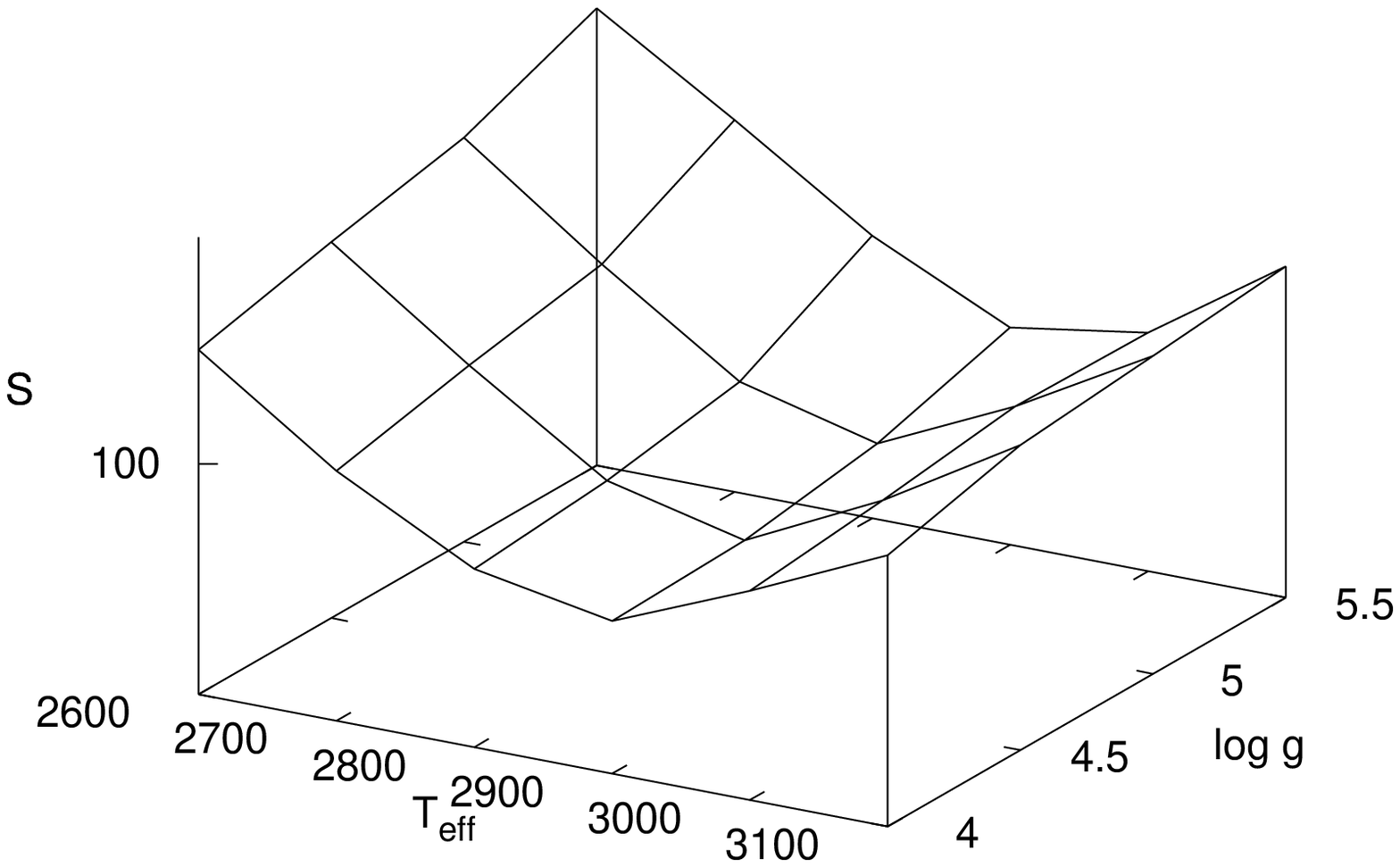}
    \includegraphics[width=0.48\linewidth, angle=0]{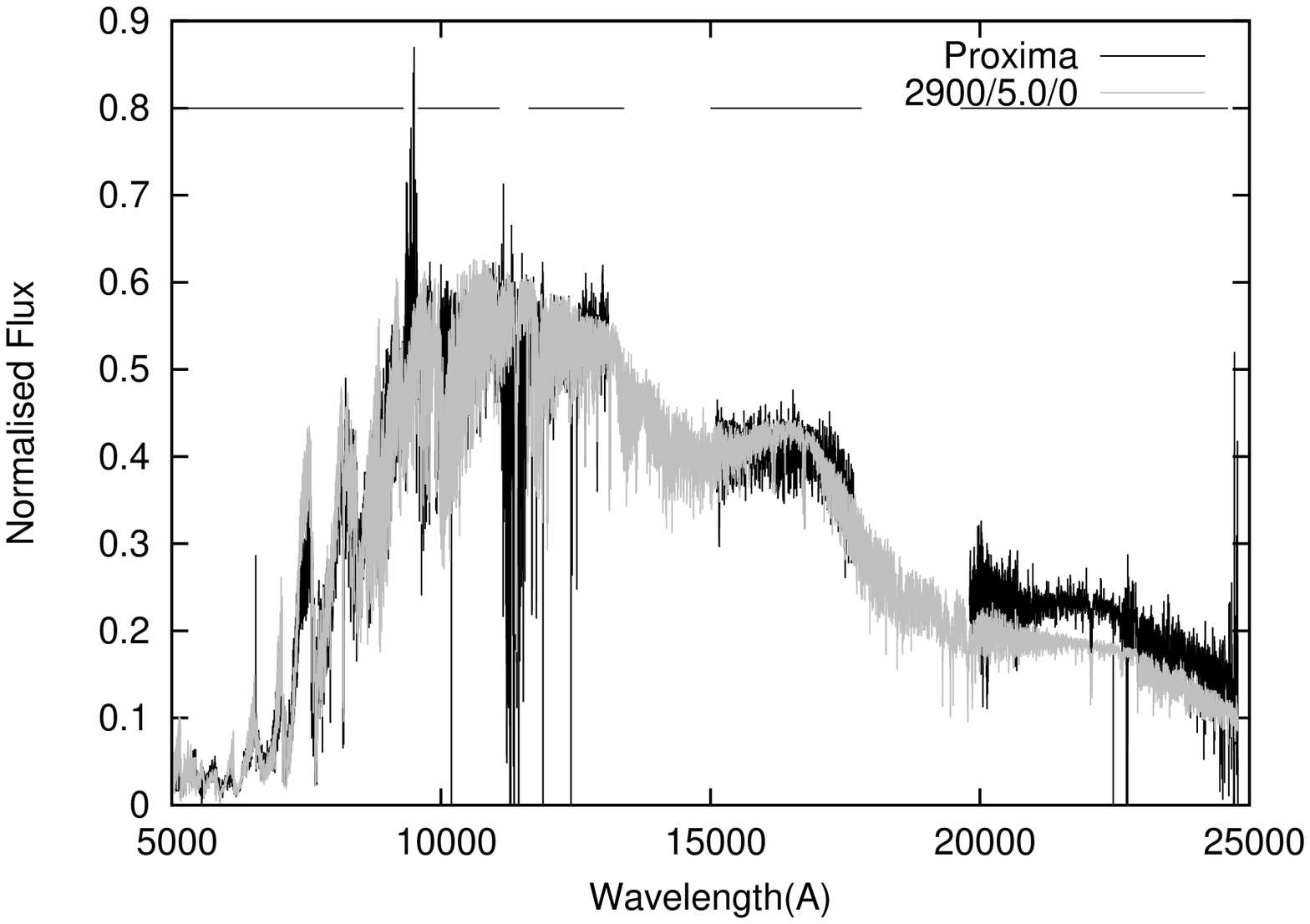}
   \caption{{\it Left}: dependence of $S$ on \Tef, log g. {\it Right}: the best fit of synthetic spectrum
   2900/5.0/0 to the observed VLT/X-shooter spectrum of Proxima. We remove here the spectral 
   regions of the strong telluric absorption in J and K bands,
   spectral ranges used for the fits are indicated by horizontal lines at $F_{\lambda}$ = 0.8.}
   \label{_1s}
\end{figure*}

\citet{rajp11} and \citet{pass16} showed that the effective temperature and gravity of normal field M5 
dwarfs are \Tef{}\,=\,2900$\pm$100 K and \logg{}\,=\,5.0$\pm$0.5\@, respectively. These values are most
likely applicable to Proxima classified as a M5.5 dwarf with solar metallicity. Such 
metallicity agrees well with the suggestion that Proxima is the third (C) component 
of the $\alpha$ Cen system, see \cite{rajp12} and references therein.

We fitted our synthetic spectra computed 
with the Phoenix model atmospheres in the effective temperature \Tef{} range [2600:3100 K] 
with incremental steps of 100K and gravity \logg{} [4.0:5.5] with step 0.5 dex  
to the observed be \vlt{} SED\@. 
 In our work we adopted the ''solar'' abundances of
\citet{_grev89}, except for iron abundance log N(Fe) = -4.5 in the scale $\sum N_i$ = 1.0. 
These abundances agree with \cite{aspl09} within accuracy $\sim$ 0.1 dex for the most elements.
Nevertheless, our abundances allow to fit the spectra of the Sun and solar like stars 
in good agreement with other authors, 
using comparative simple 1D model atmospheres, see \cite{ivan16}. 
We refer the reader to \citet{pavl14} and \citet{pavl15} 
for a review on input data and detailed explanation of the procedure employed to compute the
synthetic spectra.  Our least squares fitting procedure is described in \citet{pavl06a}.
We choose the best fit of the computed spectra to the observed spectrum for the
minimum of the $S$ function defined as:
\begin{equation}
S(f_{\rm h}, \Delta\lambda, R) =
   \sum_\lambda \left ( F_{\lambda} - f_{\rm h} * F_{\lambda}^s(\Delta\lambda,R) \right )^2  ,
\end{equation}
where $F_{\lambda}$ and $F_{\lambda}^s$ are the fluxes in the observed and computed fluxes
respectively, and $f_{\rm h}, \Delta\lambda, R$ are the normalization flux factor, shift in wavelengths
between observed and computed spectra, instrumental broadening factor $R$, respectively.
We created fits for all synthetic spectra from our grid.

In the left panel of Fig.\ \ref{_1s}, we show the dependence of $S$ computed for the range 
of adopted parameters in effective temperature \Tef{} and gravity \loggg{}. 
We carried out our minimisation procedure only for  the ''good'',  i.e. without 
notable telluric absorption and/or emission features across $\lambda\lambda$ 6650 -- 6567 \AA,
9300 -- 9575 \AA,
11082 -- 11629 \AA,
11889 -- 11894 \AA,
13393 -- 15000 \AA,
17808 -- 19638 \AA, see Fig. \ref{_1s}. 
Shorter wavelengths of $\lambda <$ 5000 \AA~ were excluded due to some problems discussed in Section \ref{_aopacity}. 
We find a clear minimum of $S$ for \Tef{}\,=\,2900\,K for all considered cases of \logg{}. 
The dependence on \logg{} is rather weak when varying the  \Teff{}.

\subsubsection{Gravity from absorption lines in NIR VLT/X-shooter spectra of Proxima.}

We showed that the fit to observed SED of Proxima suggests \Tef{}\,=\,2900\,K (Fig.\ \ref{_1s}). 
To constrain further the gravity of Proxima, we fit gravity-sensitive absorption lines present
in the optical spectrum. 
Here we draw main attention to the profiles of atomic lines.
It is worth noting that due to complicatedness of spectra of M-dwarfs 
the comparison of atomic line profiles in computed and observed spectra is not easy task. 
The atomic lines in spectra of M-dwarfs form at the background of the  haze of molecular lines
of different intensity. The molecular features/blends cannot be fitted as good as
atomic lines. Nevertheless, comparison of observed and computed profiles
of atomic lines allows us to constrain appropriate input parameters.
In particular, we focused here our efforts on the resonance lines of 
potassium at 7664.9 and 7698.96\AA{} as well as the subordinate triplet of sodium at 
8126\AA{}, which are well-known gravity indicators. We show the fits to these lines in 
Fig.\ \ref{_atop}. We can conclude that the optical spectrum of Proxima is best reproduced with
 solar abundances and \logg{}\,=\,5.0 dex. 
Our results agrees within the
uncertainties with the \logg{}\,=\,5.5 dex derived by \citet{pass16}.

\subsubsection{CO bands}

In Fig.\ \ref{_CO} we compare the observed spectrum of Proxima with the theoretical spectrum 
using the $\Delta v$\,=\,2 CO bands computed for the 2900/5.0/0 model atmosphere.
We can see that the fit of the synthetic CO bands to the observed data is a good diagnostic 
to infer the physical conditions in M dwarf atmospheres, see \cite{pavl02c}.
 In particular we find that the synthetic spectra with the proper 
effective temperature  match reasonably well the observations and agree 
with the temperatures derived by empirical methods. We see a rather marginal 
response to a presence of outer hot atmospheric layers, i.e. of a 
chromosphere, because CO is a very stable molecule of large 
dissociation potential (D$_0$ = 11.105 eV). In general, the CO bands are seen in absorption. 
At the resolution of our observations we may conclude also that our $^{12}$C/$^{13}$C ratio is consistent with the solar
because the $^{13}$CO bands are weak or even absent in the observed spectrum (Fig.\ref{_CO}). 

\subsubsection{Abundances in the Proxima atmosphere}
\label{_abundss}

\begin{figure*}
\begin{floatrow}
\ffigbox{
\includegraphics[width=\columnwidth, angle=0]{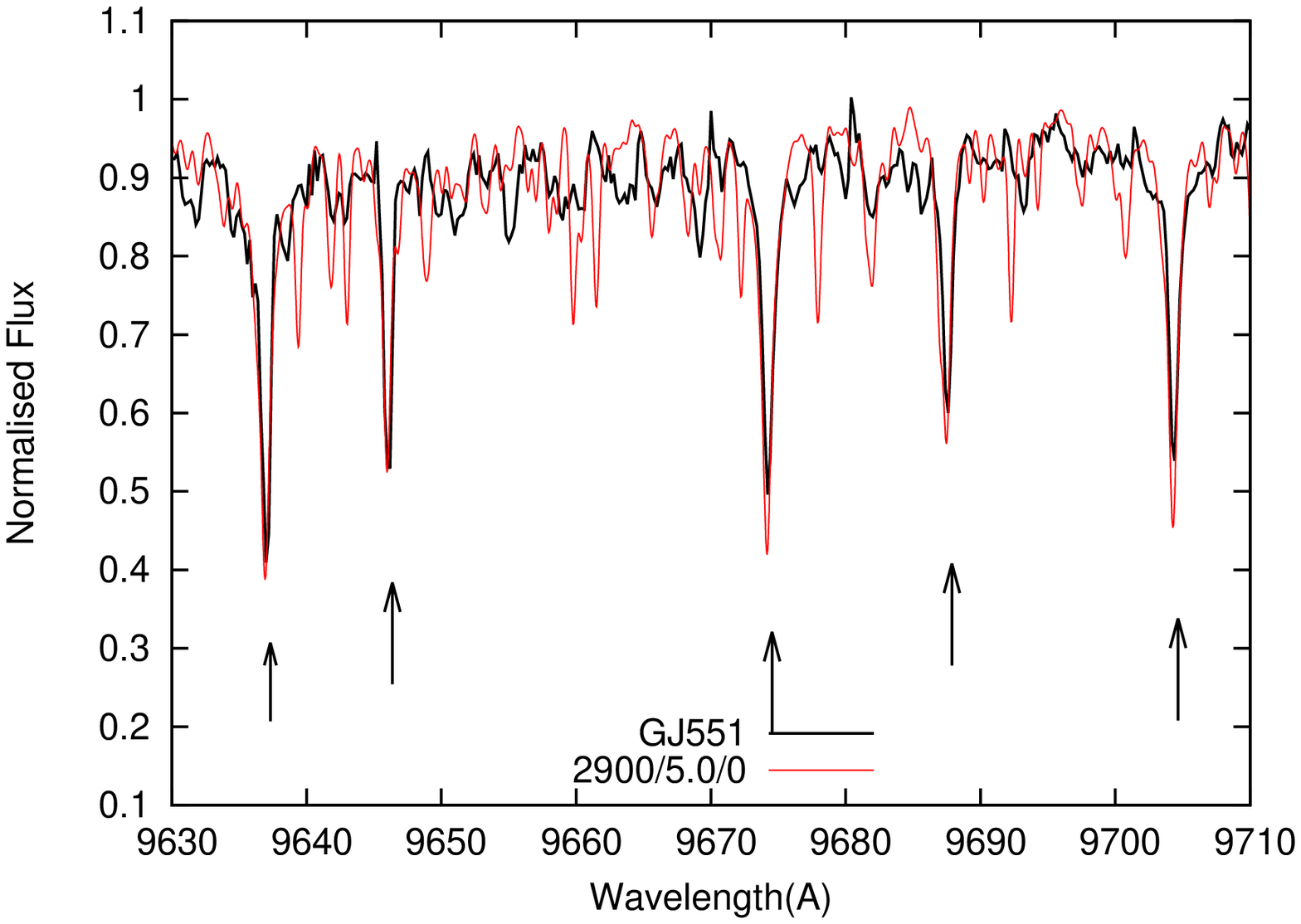}

}{
\caption[]{Synthetic fit to Ti I lines marked by arrows in the observed VLT/X-shooter spectrum.}\label{_ti96}
}
\capbtabbox{
\begin{tabular}{@{\hspace{0mm}}c c c @{\hspace{0mm}}l}
 \hline
$\lambda$   & $gf$    & E" (eV)&  log N(Ti) \\
 \hline
           &          &        &   \\
 \hline
  9638.31  & 2.44E-01 & 0.848 &  -7.05\,$\pm$\, 0.15 \\
 9647.37   & 3.68E-02 & 0.818 &  -7.05\,$\pm$\, 0.15     \\
 9675.54   & 1.57E-01 & 0.836 &  -7.35\,$\pm$\, 0.15  \\
 9688.87   & 2.45E-02 & 0.813 &  -7.35\,$\pm$\, 0.15  \\
 9705.66   & 9.79E-02 & 0.826 &  -7.20\,$\pm$\, 0.15      \\

\end{tabular}

}{%
\caption{Abundances of Ti{\small{I}} obtained from the fits to lines shown in Fig. \protect\ref{_ti96}.
 }
\label{_ti96t}
}
\end{floatrow}
\end{figure*}

In the previous sections we analysed the saturated lines of atoms and molecules, which, by 
definition, show a rather marginal dependence on the changes of abundances. We can assume
here that the abundances of the alkali elements shown in Fig.\ \ref{_atop} do not differ much 
from the solar case. In the \vlt{} spectra 
we observe also absorption lines of other elements at the background of 
the local pseudo-continuum formed by the molecular bands across the 
optical and NIR spectral ranges.  
Therefore, to make our abundance analysis more reliable, we employ a few atomic 
lines of intermediate strength present 
in the red part of the optical spectrum of Proxima where molecular absorption is weaker. We discuss
these lines below. 

{\it Titanium.} Lines of Ti{\small{I}} are more numerous than Fe{\small{I}} in the spectrum of
Proxima due to the lower potential of ionisation of titanium. In  Fig.\ \ref{_ti96} we show the fit with 
the synthetic spectrum computed for the 2900/5.0 model atmosphere with the solar 
$\log$ N(Ti)\,=\,$-$7.05 \citep{grev98}. 
In Table \ref{_ti96t} we list the derived abundances. We find a Ti abundance of 
 log N(Ti)=-7.20\,$\pm$\, 0.15 from an average of five Ti{\small{I}} lines, suggesting a weak 
metal deficient atmosphere for Proxima\@. However, we caution this abundance estimate 
because we see that most of the observed absorption lines are weaker than the lines in
the theoretical spectrum computed at solar abundance.

\begin{figure*}
  \centering
  \includegraphics[width=0.32\linewidth, angle=0]{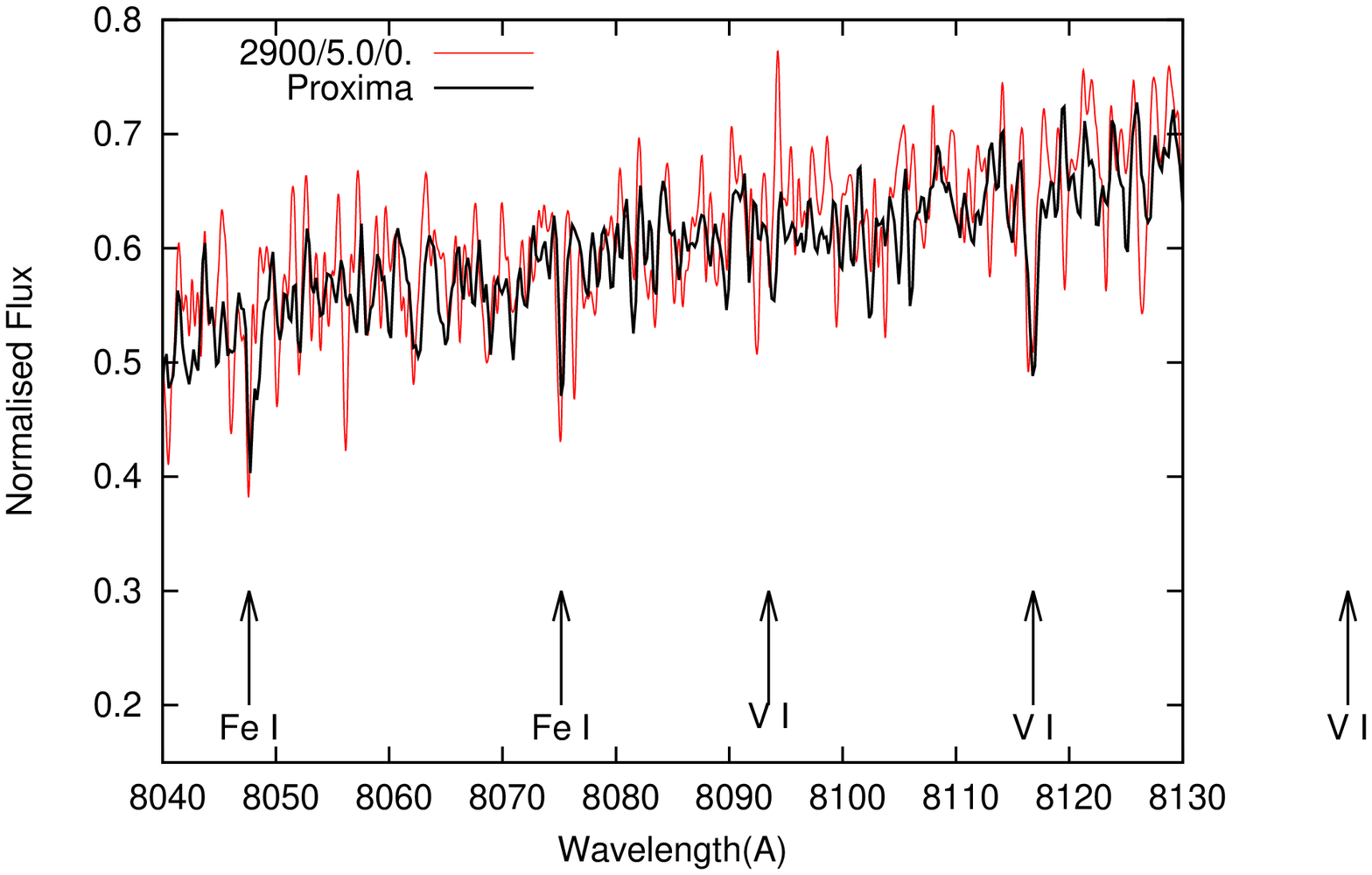}
  \includegraphics[width=0.32\linewidth, angle=0]{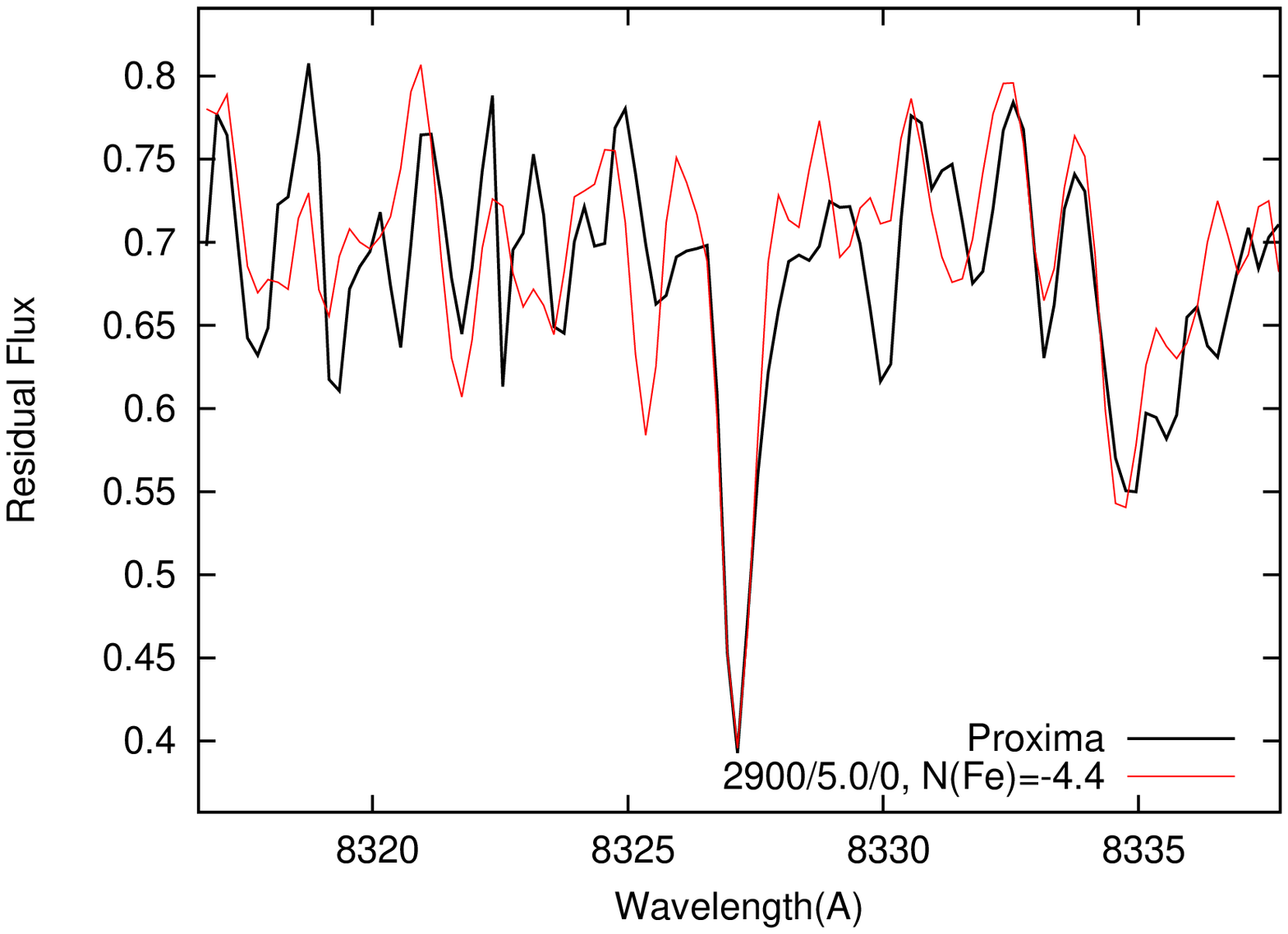}
  \caption{{\it Left}: fits to the observed Fe{\small{I}} and V{\small{I}} lines with the
  2900/5.0/0 model atmosphere.
 {\it Right}: fit to the observed Fe{\small{I}} line at 8327.06\AA{} with $\log$ N(Fe)\,=\,$-$4.4}
   \label{_Fe1}
\end{figure*}

{\it Iron.} Iron lines are numerous in the spectrum of Proxima but not as intense as the
Ti{\small{I}} lines. Although weak lines are more affected by the uncertainties, the theoretical 
and observed spectra agree qualitatively well as shown in the left panel of Fig.\ \ref{_Fe1}. 
On the right panel of Fig.\ \ref{_Fe1} we display the fit to the observed profile of the
intense  Fe{\small{I}} line at  8327.06\AA{} ($gf$ = 0.02985), showed also in Fig.\ \ref{_atop}. 
We obtained good quantitative agreement for $\log$ N(Fe)\,=\,$-$4.4 dex, similar to the
 the solar iron abundance within $\pm$0.2 dex. 

\begin{figure*}
  \centering
  \includegraphics[width=0.32\linewidth, angle=0]{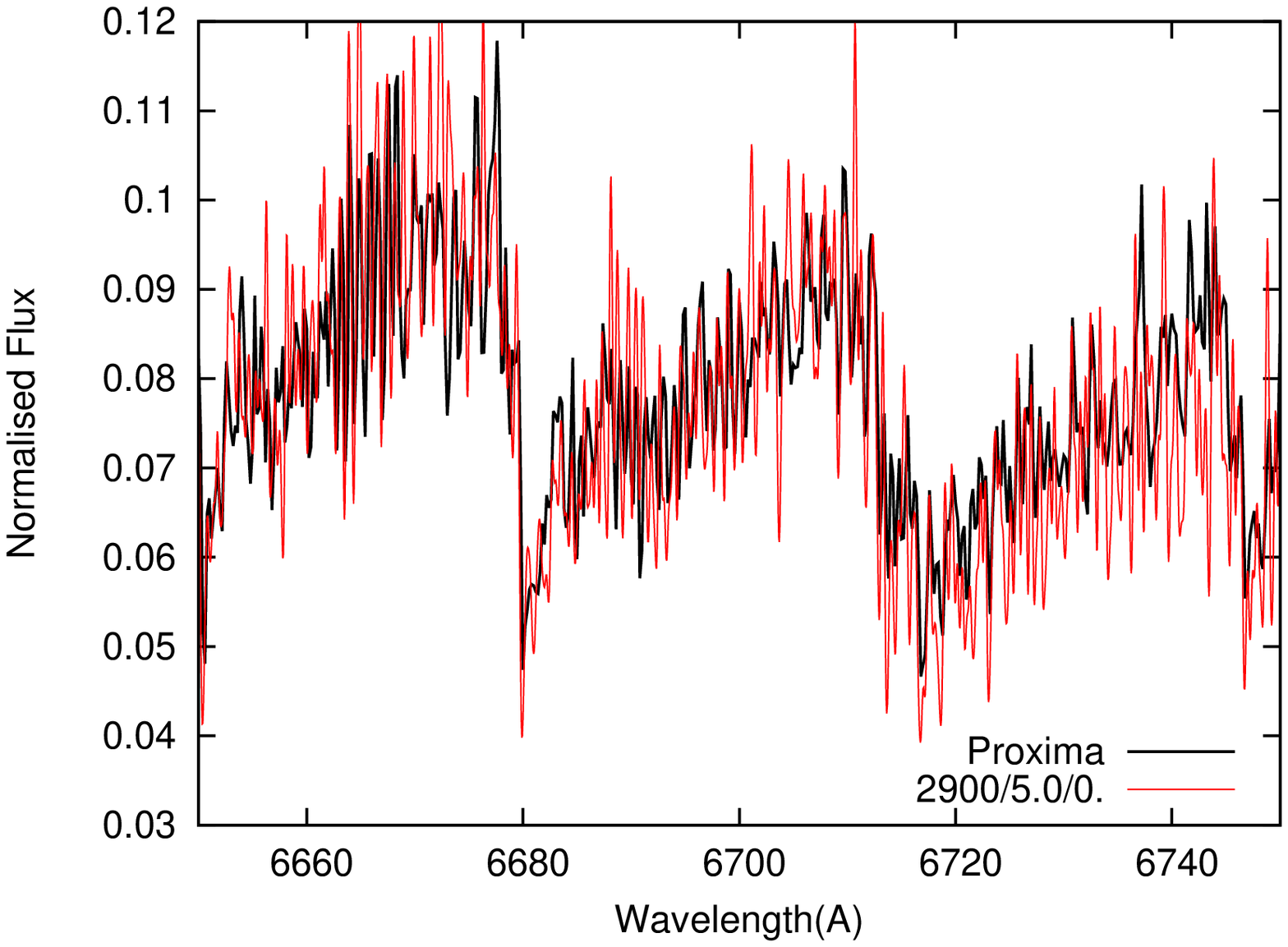}
  \includegraphics[width=0.32\linewidth, angle=0]{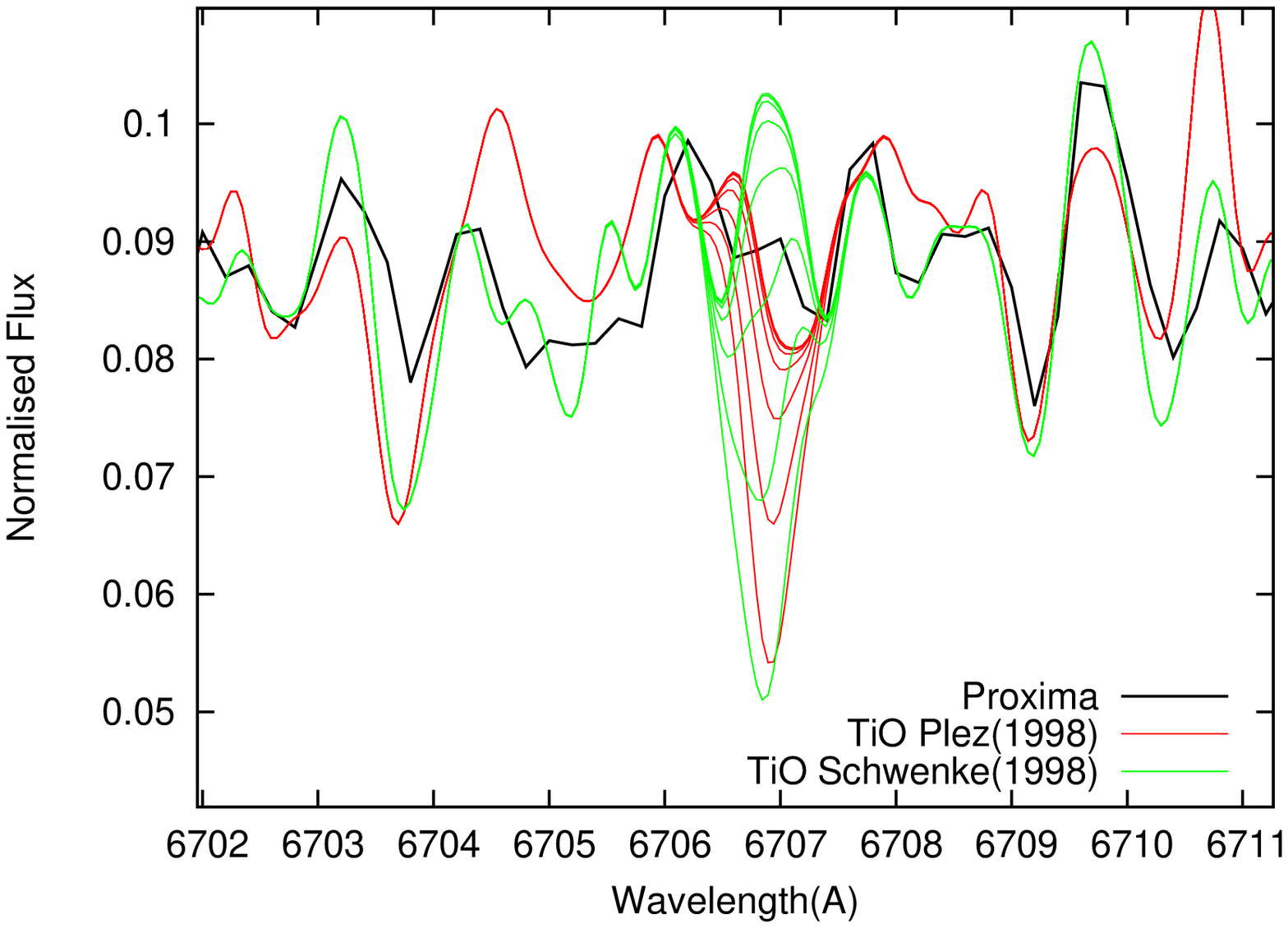}
 \caption{{\it Left}: fit to observed fluxes observed across 6708 \AA.
{\it Right}: fits of synthetic spectra computed line lists of \protect\cite{schw98} and 
\protect\cite{plez98} for different lithium abundances to 
the observed 6708\AA{} blend, lithium abundances from the bottom: -11.04, -11.54, -12.04.... 
Model atmosphere 2900/5.0/0.}
   \label{_tioLi}
\end{figure*}

{\it Lithium.} The absorption bands of the TiO molecule govern the spectra of M dwarfs 
around the Li resonance doublet at 6707.8\AA{}. The Li doublet is not seen in the spectrum 
of Proxima as expected for a  fully convective old low-mas star. 
We compare the observed by HARPS and synthetic spectra computed 
with the 2900/5.0/0.0 model atmosphere and the line lists from \citet{plez98} and \citet{schw98}.
Generally speaking, Schwenke's TiO line list allows better to reproduce 
the shape of the blends across spectral range containing Li doublet. However, in our case
we can only place an upper limit to the lithium abundance in the atmosphere of Proxima at 
$\log$ N(Li)\,=\,$-$12.04 dex (right panel in Fig.\ \ref{_tioLi}).  
 
\subsubsection{Atomic and molecular absorption spectra in the 3800--4200\AA{} region}
\label{_aopacity}

Absorption lines of neutral species in the blue spectra of M-dwarfs are expected to be much stronger
than in the solar case 
due to the lower temperatures and higher pressures in the regions of the absorption line formation. 
 Moreover, we may expect resonance lines of neutral metals to appear due to 
the changes of their ionisation equilibrium in this low temperature regime.

To verify our treatment of the pressure broadening we compute two spectral regions in the solar
spectrum containing rather strong enough lines. 
We follow the procedure 
described in \cite{_pavl95} in our computations. The profile of the absorption line is 
described by a Voigt function $H(a,v)$ and the damping broadening parameter $a$ changes 
with depth in stellar 
atmosphere. We computed the synthetic spectra with the VALD3 line list 
\citep{_ryab15, ryab15a} 
for the 5777/4.44/0 solar model atmosphere \citep{_pavl03} with a micro-turbulent velocity of
\Vm =1 \kmps{} and wavelength steps of 0.025\AA{}.  In Fig.\ \ref{_sun} we observe a good agreement between the profiles of 
strong atomic lines in the observed spectrum of the Sun as a star and the computed spectrum.
(Fig.\ \ref{_sun}).

\begin{figure*}
\begin{floatrow}
\ffigbox{%
\includegraphics[width=\columnwidth, angle=0]{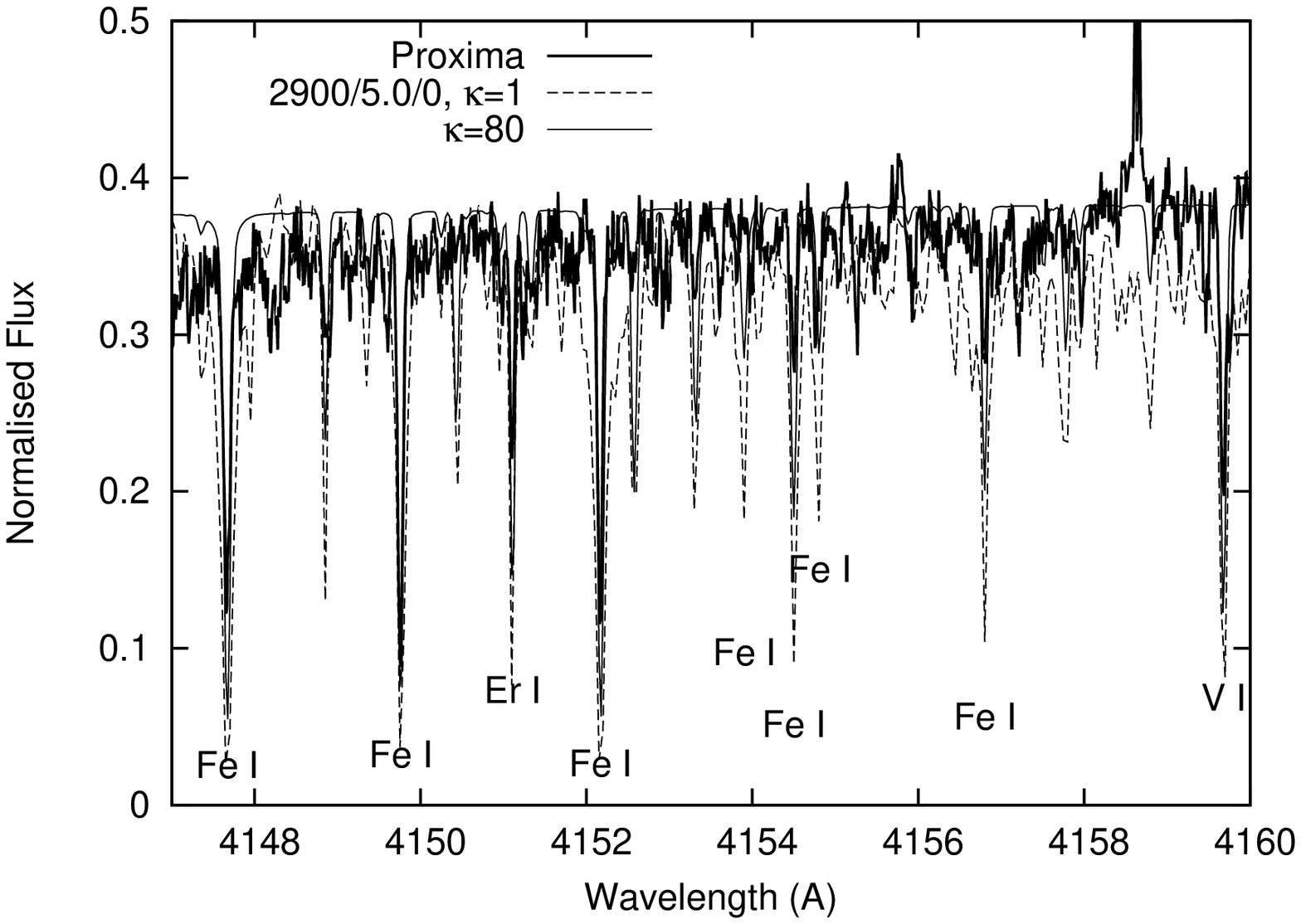}

}{
\caption[]{Identification of absorption lines in the arbitrary selected part in the observed HARPS
spectrum of Proxima\@.}
\label{_idents}
}
\capbtabbox{
\begin{tabular}{@{\hspace{0mm}}c c c c@{\hspace{0mm}}l}
 \hline
$\lambda$   & Element  &   $gf$      & E" (eV)&  Visibility \\
 \hline
           &          &         &      &   \\
 \hline
  4147.67  &    Fe{\small{I}} &7.87E-03 & 1.485 & Yes \\
  4149.76  &    Fe{\small{I}} &4.93E-06 & 0.052 & Yes \\
  4151.11  &    Er I &2.73E+00 & 0.000 & Yes \\
  4152.17  &    Fe{\small{I}} &5.86E-04 & 0.958 & Yes \\
  4153.90  &    Fe{\small{I}} &4.77E-01 & 3.397 & No \\
  4154.50  &    Fe{\small{I}} &2.05E-01 & 2.832 & No \\
  4154.81  &    Fe{\small{I}} &3.98E-01 & 3.368 & No \\
  4156.80  &    Fe{\small{I}} &1.55E-01 & 2.832 & No \\
  4159.68 &     V I& 1.86E-02&  0.287& Yes \\
\end{tabular}

}{
\caption{Identification of lines in  the HARPS spectrum of Proxima shown in the left panel of 
Fig.\ref{_idents}} \label{_identt}
}
\end{floatrow}
\end{figure*}

We computed the synthetic spectrum of Proxima across the 3800--4200\AA{} wavelength
range. Molecular absorption is weak or absent at these wavelengths, implying that we can
see deeper layers of the photosphere of Proxima\@. However, the comparison of the 
intensities of observed absorption lines compared with the computations reveals reveals some 
problems here:
\begin{itemize}
\item
A simple analysis of the lines listed in Table \ref{_identt} shows that only lines with 
low excitation energy (E" $<$ 2 eV) are seen in the observed photospheric spectrum of Proxima
(Fig.\ \ref{_idents}). We see that the lines with higher excitation energies are weaker (or even 
absent) in the observed spectrum. 
\item The strongest atomic lines in the observed spectrum are much stronger in the 
synthetic spectra computed in the framework of the classical approach. In other words, 
damping pressure effects are more pronounced in the computed spectrum where atomic 
lines have more extended wings.
\item  Our numerical experiments show that changes of effective temperature by $\pm$200\,K
or $\log$(g) by $-$0.5 to $-$1.0 do not improve the fit. We cannot reduce the intensities of 
saturated lines by reducing the associated abundances because we know from the Section 
\ref{_abundss} that the metallicity of Proxima is near solar.
\end{itemize}

\begin{figure*}
\centering
\includegraphics[width=75mm, angle=0]{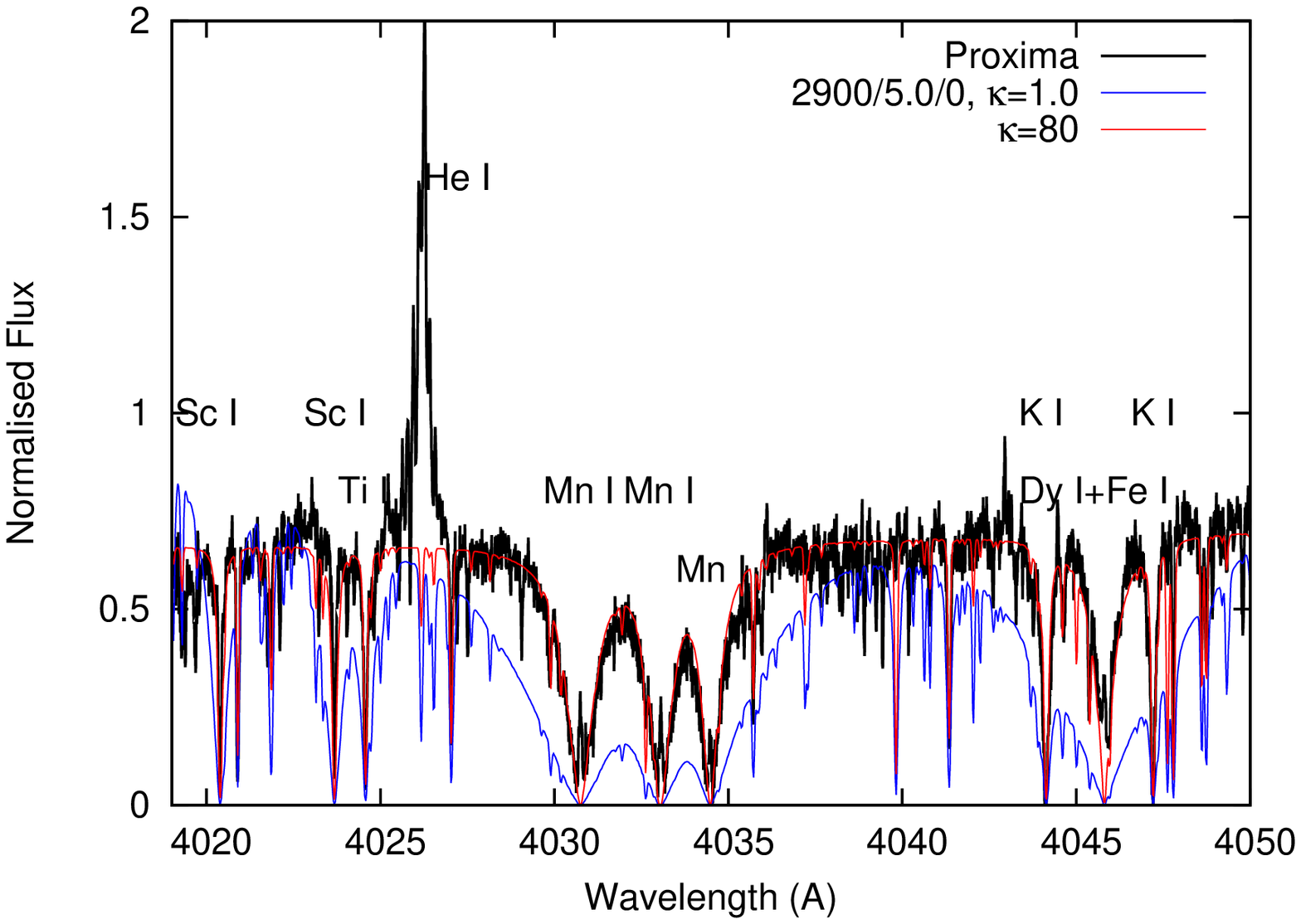}
\includegraphics[width=75mm, angle=0]{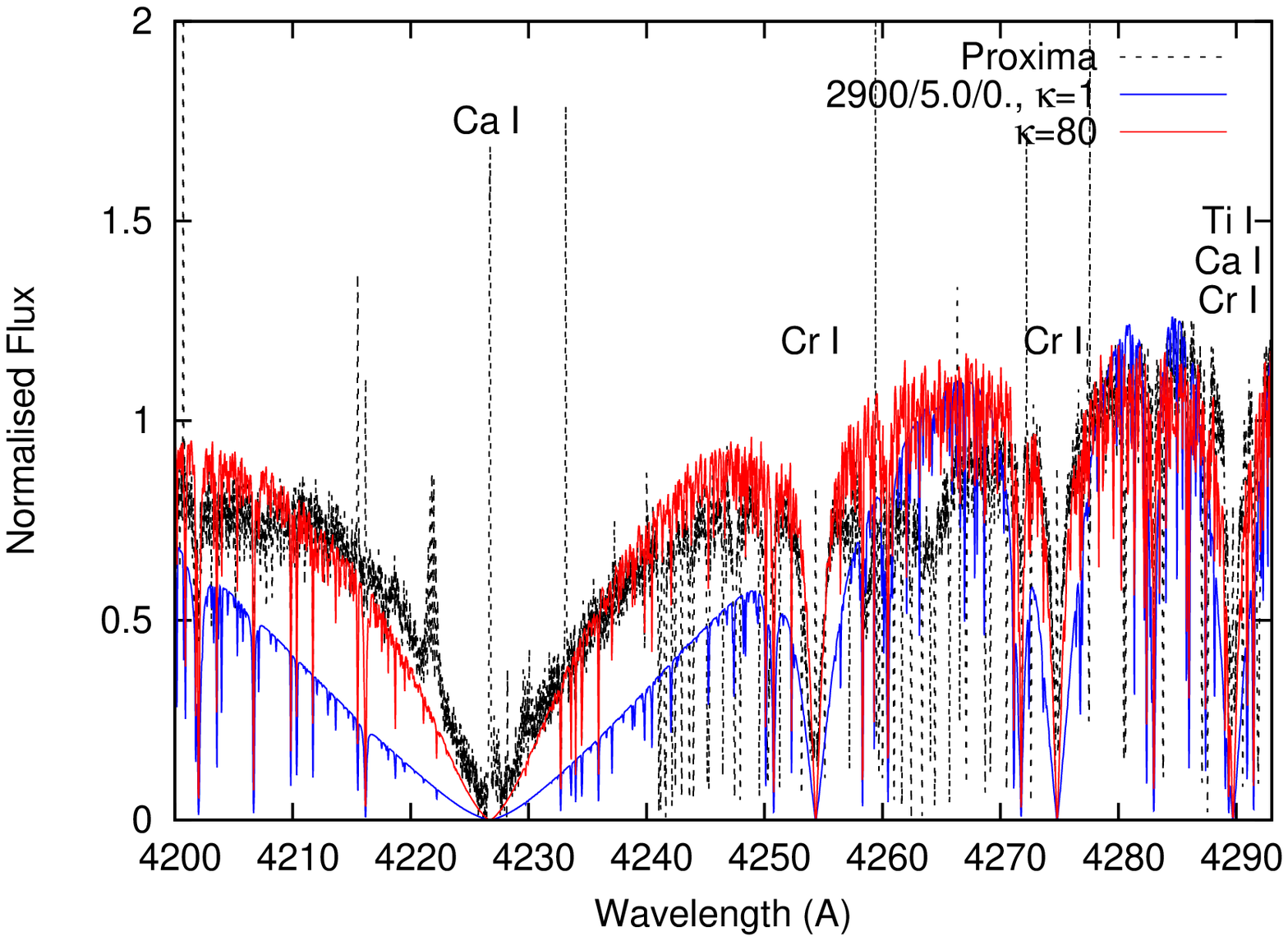}
\caption{The spectrum of Proxima computed for the model with enhanced 
continuum opacity across the blue spectral region.}
\label{_adop}
\end{figure*}

We can explain the differences by enhancing the continuum opacity, as shown in 
Fig.\ \ref{_adop} where we compare the observed spectrum with the newly computed one.
To reduce the strength of resonance lines we could move the line forming region 
into lower pressure regions of the atmosphere. In this paper, we use a simple approach
suggesting $\chi^c_{\nu} = \chi^{c0}_{\nu} \times \kappa$, where $\chi^{c0}_{\nu}, \kappa$ 
are the conventional opacity and adjusting parameter, respectively.
Enhancing the continuum opacity across blue spectral range shifts the line forming 
regions upwards, i.e.\ to layers of lower pressure. As a consequence, the strongest lines in 
the computed spectrum show weaker wings, in better 
agreement with observations. We obtain satisfactory fits after implementing additional
continuum opacity in the blue, and can explain the lack of lines of higher 
excitation potentials, as that case, the photosphere of the star moves upwards 
into layers of the atmosphere 
with lower temperature, where only lines of low excitation potential can form. Absorption lines 
of lower excitation energies form above the ''new'' photosphere, making them less sensitive 
to these changes (Fig.\ \ref{_idents}).

\subsection{Emission lines}

 The intensity of the emission features visible in the spectra changes during the different activity states of the star. 
To analyse the difference between the typical activity level and the more quiet states we created two average spectra 
representing the typical active state (S) and the quiet state (QC), see section ~\ref{harps_s_qc}. 
The 'QC' spectrum shows lower intensity for most of the measured emission lines, specifically
the intensity of H$\alpha$ here does not exceed 75\% of the maximum emission achieved during strong flares.

\subsubsection{Hydrogen Balmer lines}

All Balmer lines in the spectrum of \pr{} are observed in emission. They show strong variability 
in time with variations up to a factor of 10\@ in intensity. 
As a complementary information, we provide
movie\footnote{ftp://ftp.mao.kiev.ua/pub/yp/2017/p/halpha.avi}  
with the time variability of the \Ha{} in the series of data obtained with HARPS.  
We measured the pseudo-equivalent widths (pEW) of the emission lines of the full Balmer
serie in the VLT/X-shooter spectrum with the task {\tt{splot}} under IRAF, which we report
in Table \ref{tab_ProxCen:table_pEW_XSH}.
Interestingly, measurements of pEW of Balmer lines in the averaged HARPS 'S' spectrum
provides values of the same order than those reported in the Table.

%
%
\begin{table}
 \centering
 \caption[]{Central wavelengths ($\lambda_{c}$) and pseudo-equivalent widths measured
 on the VLT/X-shooter spectrum and expressed in\AA{} for the observed Balmer serie}\label{tab_ProxCen:table_pEW_XSH}
 \begin{tabular}{@{\hspace{0mm}}c c c c@{\hspace{0mm}}}
 \hline
Line    & j-i    &   $\lambda_{c}$      & pEW \cr
 \hline
        &    &     \AA{}       &\AA{} \cr
 \hline
H$\alpha$ & 3--2   & 6562.797$\pm$0.02 &  2.5$\pm$0.5 \cr
H$\beta$  & 4--2 & 4861.323$\pm$0.02 &  5.1$\pm$0.2 \cr
H$\gamma$ & 5--2 & 4340.462$\pm$0.02 &  6.5$\pm$0.5 \cr
H$\delta$ & 6--2 & 4101.734$\pm$0.02 &  9.9$\pm$0.2 \cr
H$\epsilon$&7--2 & 3970.072$\pm$0.02 &  5.4$\pm$1.0 \cr
H$\zeta$  & 8--2 & 3889.048$\pm$0.02 &  9.4$\pm$0.8 \cr
H$\eta$  & 9--2 & 3835.384$\pm$0.02 &  3.9$\pm$0.1 \cr
H$\theta$ & 10--2 & 3797.898$\pm$0.02 &  4.7$\pm$0.4 \cr
 \hline
  \end{tabular}
\end{table}

\begin{figure*}
\centering
\includegraphics[width=0.48\columnwidth, angle=0]{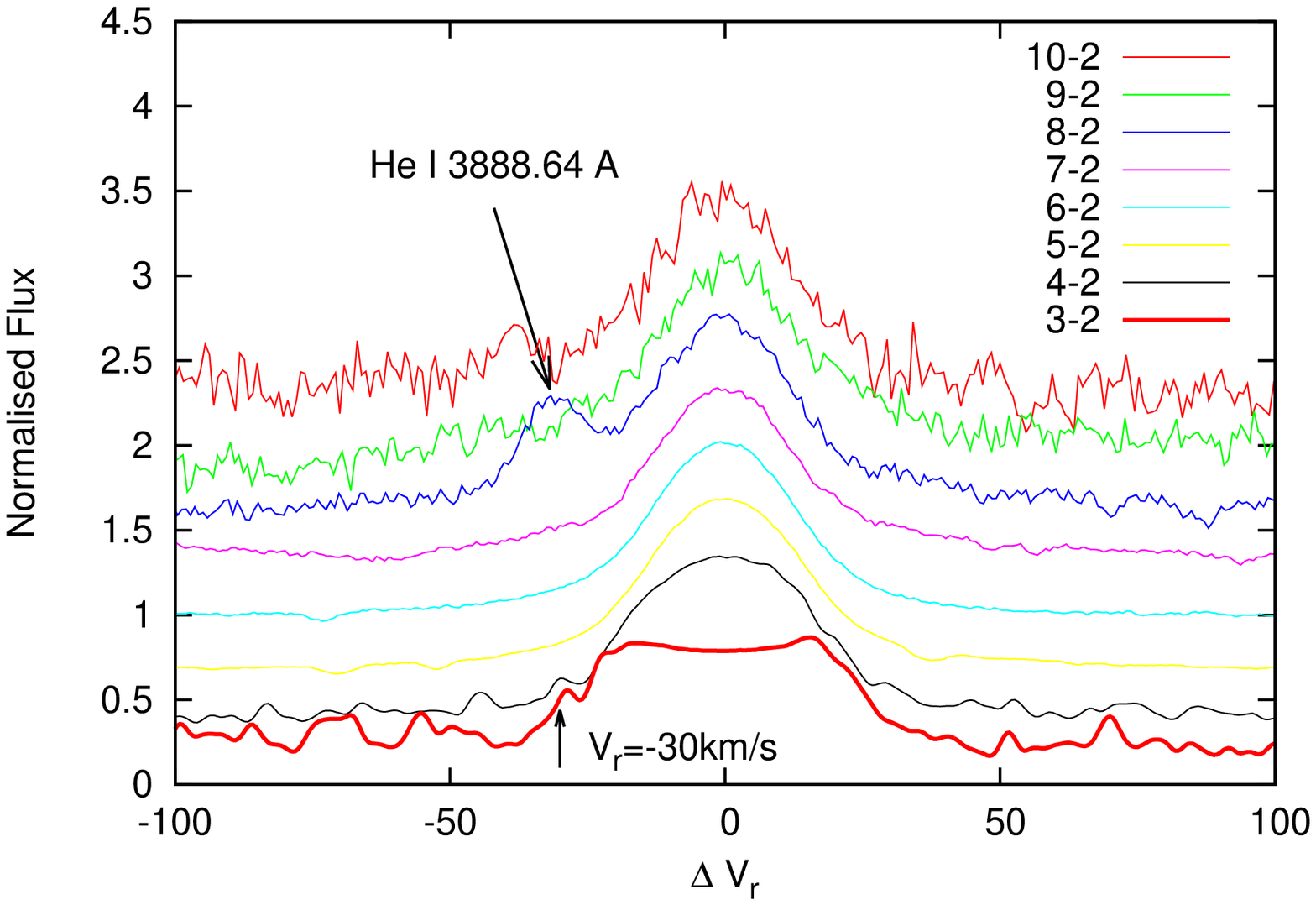}
\includegraphics[width=0.48\columnwidth, angle=0]{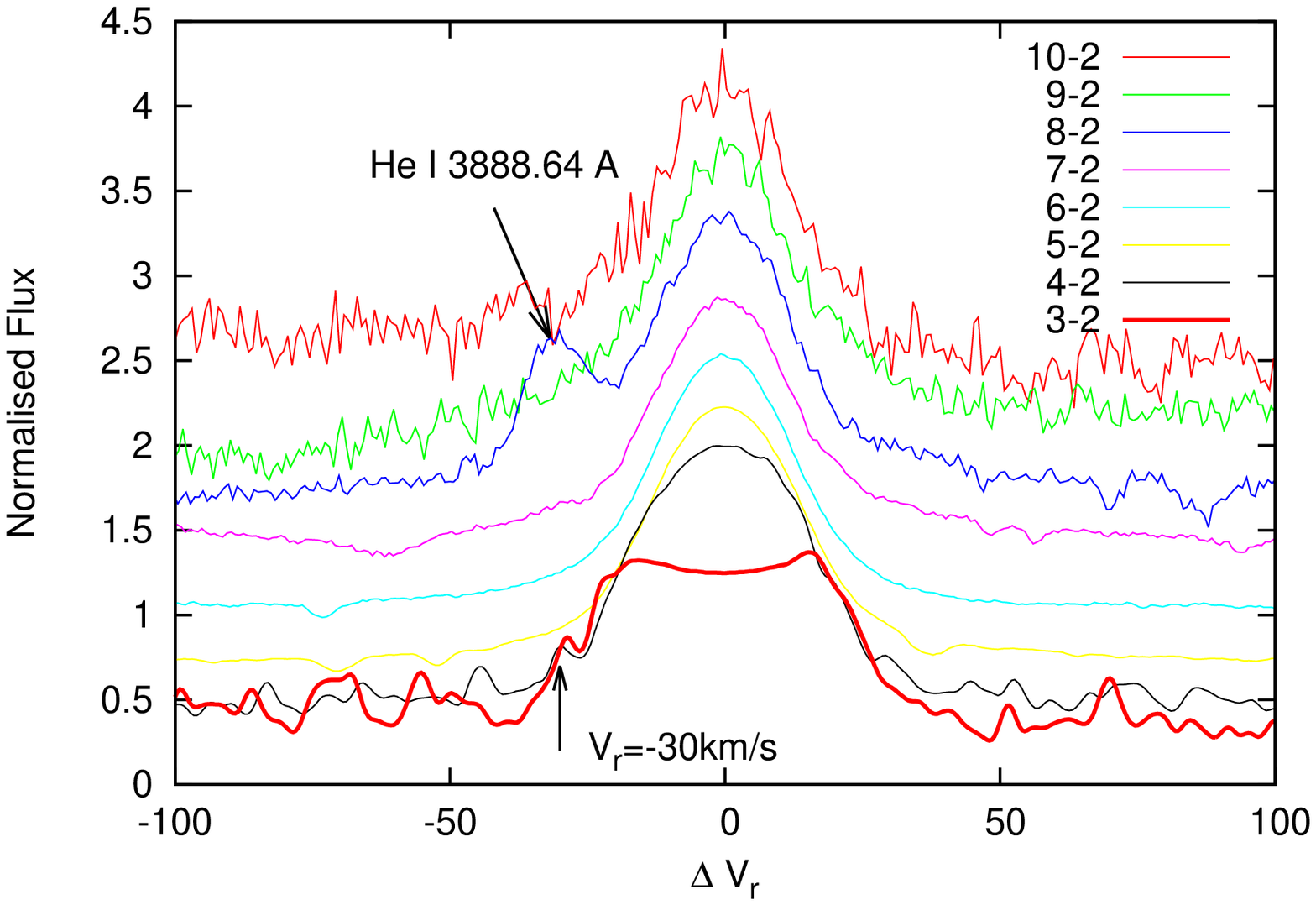}
\caption{Profiles of Balmer lines shown in the flux
normalised to 1.0 in the center of lines vs. radial velocity
obtained in quiet mode (left) and flare mode (right). 
To simplify the plot profiles are shifted in vertical scale, \Ha{} is shown by solid line.
The arrows mark the positions of the \Vr = -30 \kmps{} component feature and He I 3888.64\AA{} line.}
\label{_s}
\end{figure*}

 In Fig.\ \ref{_s} we show the intensity profiles of the H${\alpha}$ line vs Doppler velocity
 (\Vr{}\,=\,$\Delta \lambda /\lambda \times c$), where $c$ is the speed of light.
 We show the profiles for both the quiet (left; QC) and flare (right; S) states.
 In our spectra we see the full Balmer serie, from the $H_{\alpha}$ line to $H_{\theta}$, 
 corresponding to the 3--2 and 10--2 transitions of the hydrogen atom, respectively.
Comparison of their emission profiles provides at least a few important results:

\begin{itemize}
\item All Balmer lines show strong variability responding to the
temporal changes of flare activity.
\item Self-absorption in the core of the \Ha{} line is observed practically for all
stages of activity of Proxima (see movie$^3$ in supplementary data). We interpret this
phenomenon as evidence for the existence of comparatively cool matter outside the flare region.
The peak of the line on  the red side is higher than on the the blue, most likely
due to the outward motion of the neutral hydrogen as discussed in \citet{fuhr11}.
\item The positions of lines do not change during flare events. We interpret this
fact as evidence for a quasi-stationary state of the region where Balmer emission lines form.
\item The profiles of all Balmer lines in emission in the intensity vs. doppler 
velocity parameter space have the same FHWM despite of the differences in intensity.
The similarity in the profiles of emission lines displayed in Fig.\ \ref{_s} is an evidence 
that they form in the same region of the atmosphere of \pr{} heated by flares.
\item Some lines in the Fig.\ \ref{_s} show a well-pronounced peak at $V_{r}$\,=\,-- 30 \kmps,
which we interpret as a indication of an hot stellar wind moving outwards from the star.
This peak is well detected in 
 in $H_{\alpha}$ and $H_{\beta}$,
 and seen as a wide feature in the blue wings of $H_{\gamma}$ - $H_{\epsilon}$.
\item In the $H_{\zeta}$ profile, we again see a well-pronounced emission feature at $V_{r}$\,=\,--30 \kmps,
however, in this case we identify this as He I line at 3889.64 \AA, see section \ref{_het}.
\item The component shifted blue-ward does not change much with the activity phase,
suggesting that it probably forms in the hot ionized plasma flow (stellar wind) far enough
from the flare region of the star.
\item When the flare activity rises, the increased emission in \Ha{} covers the line in the
blue wing (see movie$^3$ in supplementary material). This detail is also seen in Fig.\ 14 of 
\citet{fuhr11} at lower level activity times.
\end{itemize}

\begin{figure}
\centering
\includegraphics[width=0.8\columnwidth, angle=0]{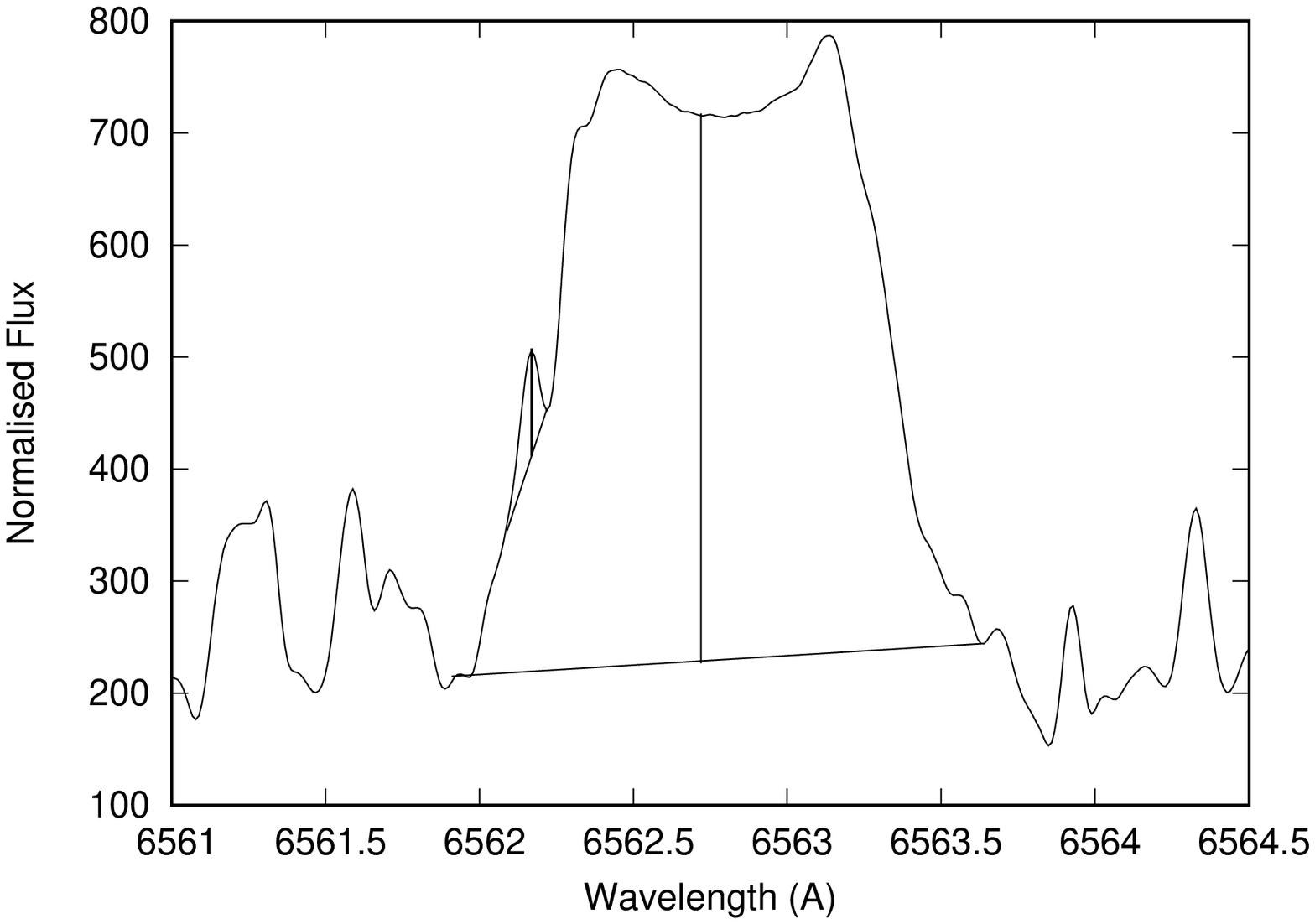}
\caption{Scheme of the measurements of  the pseudo-equivalent widths of \Ha{} and  \Vr = -30 \kmps{} feature in HARPS spectra of 
\pr.}
\label{_splot}
\end{figure}

 Photospheric spectrum as well as fluxes show rather marginal responses on the flare activity.
In the following we use flux ratios measured in \vlt{} spectra in combination with
measurements of pseudo-equivalent widths of \Ha{} and emission feature in its blue wing in the HARPS 
 'S' and 'QC' spectra as shown in Fig. \ref{_splot}.

To determine the $H_{\alpha}$ flux of \pr{} received at 
Earth we  followed the procedure by \cite{herbig85}:
\begin{equation}
f_{H_{\alpha}}=  pEW_{H_{\alpha}}  \frac{F_{\lambda}(6563)}{F_{\lambda}(5556)} F_{\lambda}(6563,0.0) 10^{-0.4V}
\end{equation}
where the second term is the average flux ratio at the indicated wavelengths 
for the star. The third term is the \vlt{} flux at 6563\,\AA{}, received from a star of 
magnitude $V$\,=\,0 mag, assumed to 
be 3.8 10$^{-9}$ erg cm$^{-2}$ s$^{-1}$ \AA$^{-1}$.  
Despite of the noise in the X-Shooter spectrum at 5556\AA{}, 
we estimated the second term from the X-Shooter spectrum and measured a flux ratio of  
$\frac{F_{\lambda}(6563)}{F_{\lambda}(5556)}$= 3$\pm$2. 

The pEW values of 2.7\,$\pm$\, 0.1 and 2.6\,$\pm$\, 0.1\AA{} were determined in 'S' and 'QC' spectra, respectively, with
 the average value  of $pEW_{H_{\alpha}}$= 2.65 $\pm$0.1 
\AA{}. For a magnitude of $V$\,=\,11.13 for Proxima \citep{jao14a}, 
we derived an average value of $f_{H_{\alpha}}$= 1.07 $\pm$0.7 10$^{-12}$ erg  
cm$^{-2}$ s$^{-1}$.

In addition, the second term can be estimated from the $R-I$ colour, 
using the relationship found by \cite{hodgkin95}  
\begin{equation}
\frac{F_{\lambda}(6563)}{F_{\lambda}(5556)} = 2.786 (R-I)_C  - 0.932
\end{equation}
where $R-I$= 2.04\,mag for Proxima \citep{jao14a}, and 
$\frac{F_{\lambda}(6563)}{F_{\lambda}(5556)}$= 4.75$\pm$0.03, 
resulting $f_{H_{\alpha}}$=  1.7$\pm$0.1 10$^{-12}$ erg  cm$^{-2}$ s$^{-1}$.  

As a final result  we adopted the mean value of both measurements, 
resulting $f_{H_{\alpha}}$= 1.4$\pm$0.4 10$^{-12}$ erg cm$^{-2}$ s$^{-1}$.
From this value of $f_{H_{\alpha}}$, we obtained a luminosity in the 
$H_{\alpha}$  of $L_{H_{\alpha}}$ = 2.8$\pm$0.4 10$^{+26}$ erg s$^{-1}$ 
adopting the distance of 1.30 pc for Proxima from \citet{jao14a}) 
and a $L_{H_{\alpha}}$/$L_{bol}$ = 4.5 $\pm$0.4 10$^{-5}$ adopting the 
bolometric luminosity of 6 10$^{+30}$ erg s$^{-1}$ from \citet{fuhr11}.

We determine \pew of the emission feature seen in the blue emission wing of \Ha{} at 
$V_{r}$\,=\, -- 30 \kmps{} in the average HARPS spectrum as shown in Fig. \ref{_splot}, 
We measure in 'S' and 'QC' HARPS spectra \pew = 0.018 \AA,  
 the ratio \pew/$pEW_{H_{\alpha}}$\,=\,0.007\@.
It allows us to estimate the total energy emitted by the stellar wind in \Ha{} 
\begin{equation}
L_{H_{\alpha}}^b = 7.0 \times 10^{-3} \, \times \,2.8 \times 10^{26} =\,2.0\times10^{+24}.
\end{equation}

Assuming complete 
ionisation of hydrogen in the emitting region we determine the number of 
emitting $H_{\alpha}$ atoms, i.e lower limit of mass loss:
\begin{equation}
 \dot{M} = L_{H_{\alpha}}^b / h\nu * m_p *N_{sec} = 3.7 * 10^{19} g/yr = 1.8*10^{-14} M_{\odot}/yr, 
\end{equation}
where $m_p$ and $N_{sec}$ are the mass of H I atom and
the number of sec in 1 year.

\subsubsection{He{\small{I}} line at 4026.19 and 3888.64 \AA}
\label{_hee}
The He{\small{I}} line at 4026.19\AA{} was identified by \citet{fuhr11}, see their Fig.\ 13 and
Table \ref{_het}. We observe strong variability of this line in the observed spectra.
We note that the He{\small{I}} line has the highest excitation energy (\EE = 20.97 eV (see
Table \ref{_hett}) with respect to other emission lines observed across our spectral range. 
It mostly likely formed in the outermost layers of the atmosphere heated by shock waves. 
Indeed, the He{\small{I}} cannot be formed in the same place where emissions in the cores 
of absorption lines of neutral metals form. Furthermore, the He{\small{I}}  emission line is 
broader.

\begin{figure}
\centering
\includegraphics[width=\columnwidth, angle=0]{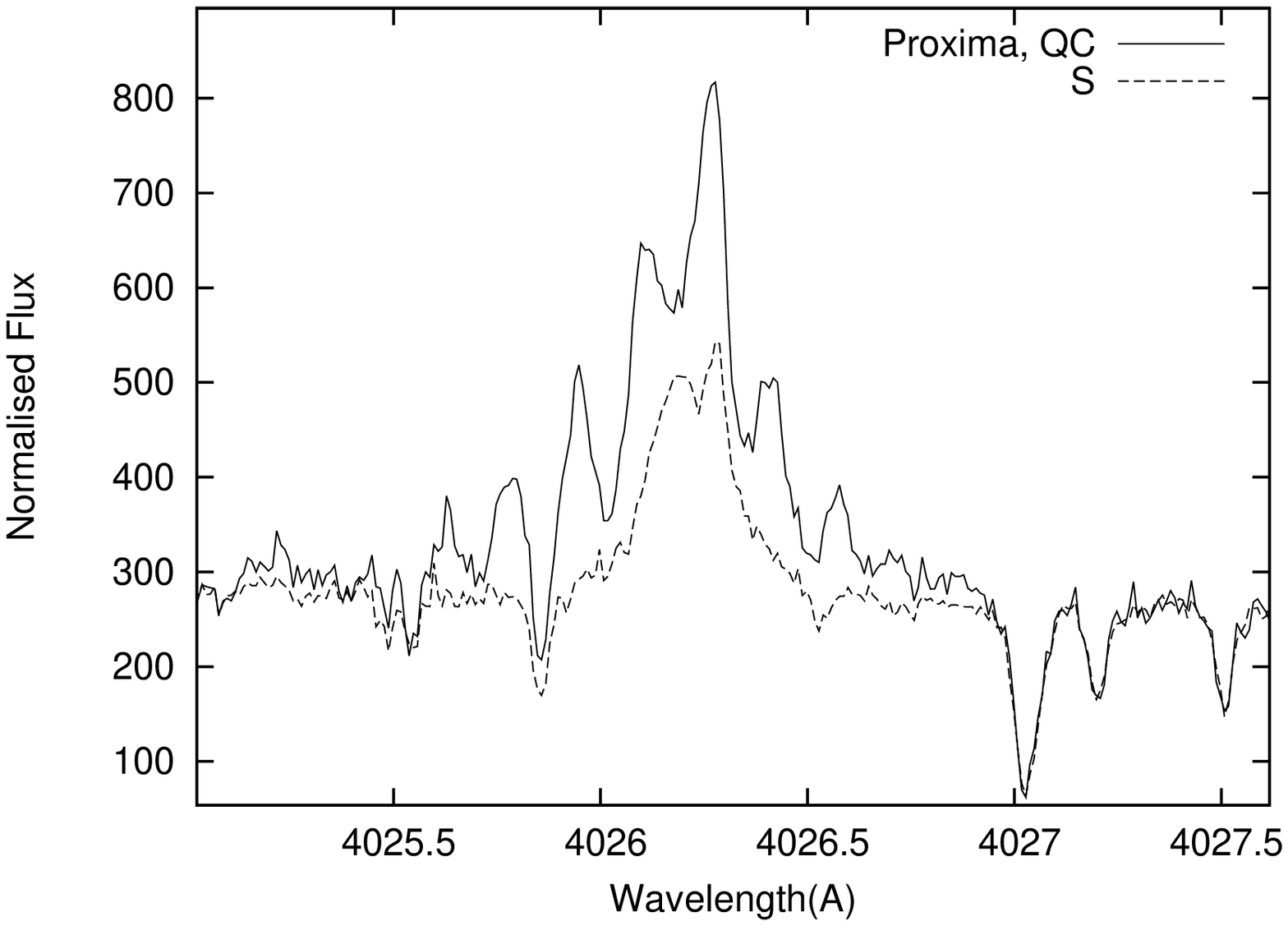}
\caption{Profiles of He{\small{I}} 4026\AA{} line observed in 'S' and 'QC' states of
flare activity}
\label{_hep}
\end{figure}

In the 'QC' dataset, the He{\small{I}}  line shows multiplet structure and looks more intensive in comparison with S
state (Fig \ref{_hep}). We suggest that the strong 
flares destroy the extended emitted region where the line is formed.
Likely, broad component seen in the 'S' spectrum can be associated with 
flare region, where, by definition, dispersion of velocities to be larger.
 In the more quiet modes,
we see a few shells moving outwards from the star represented by components equally 
shifted blue-wards and red-wards suggesting a multi-component for the line, see
Table \ref{_hett}.

Other He I line of is seen in the blue wing of hydrogen H$_{\zeta}$
line, at 3888.64 \AA. Excitation potential of this line is only a bit lower (19.82 eV) in comparison with 4026\AA{} line, 
see Table \ref{_het}. Both lines do not show any remarkable wavelength 
shift, so we may assume they form in in the same extended quasi stable hot layers. 

\begin{table}
 \centering
 \caption[]{Spectroscopic parameters of the emission He{\small{I}} lines}
 \begin{tabular}{@{\hspace{0mm}}c c c c c @{\hspace{0mm}}}
 \hline
Wavelength &   Terms  &  $j^{\prime\prime}$--$j^{\prime}$ &     \EE -- $E^{\prime}$     \\  
    (A)     &         &        &      (cm$^{-1}$)          \\
\hline
             &          &        &                              \\
  4026.18436 &      3\Po-3D &    2--1 &      169086.87 -- 193917.26 \\
  4026.18590 &       3\Po-3D &    2--2 &      169086.87 -- 193917.26 \\
  4026.18600 &      3\Po-3D &    2--3 &      169086.87 -- 193917.25 \\
  4026.19675 &     3\Po-3D &    1--1 &      169086.95 -- 193917.26 \\
  4026.19829 &        3\Po-3D &    1--2 &      169086.95 -- 193917.26 \\
  4026.35695 &       3\Po-3D &    0--1 &      169087.93 -- 193917.26 \\
             &              &         &                              \\ 
   3888.60467  &     3S-3\Po  &         1--0 &      159856.08 -   185564.96 \\
   3888.64560  &     3S-3\Po  &         1--1 &      159856.08 -   185564.69 \\
   3888.64893  &     3S-3\Po  &         1--2 &      159856.08 -   185564.67 \\
\end{tabular}
\label{_het}
\end{table}

\begin{table}
 \centering
 \caption[]{Observed components of the emission He{\small{I}} 4026\AA{} line}
 \begin{tabular}{@{\hspace{0mm}}c c c  @{\hspace{0mm}}}
 \hline
NN & $\Delta\lambda$ &   \Vr  \\ 
   &  (\AA)          & (\kmps)         \\
\hline
  1 & $-$0.560  & $-$41.73   \\
  2 & $-$0.420  & $-$31.29   \\
  3 & $-$0.240  &  $-$17.88  \\
  4 & $-$0.080  &  $-$4.59  \\
  5 & 0.090     &   6.71   \\
  6 & 0.220     &  16.39   \\
  7 & 0.380     &  28.32   \\
\end{tabular}
\label{_hett}
\end{table}

\subsection{Emission cores of resonance lines of atoms and ions}
\subsubsection{H and K of Ca II}

Ca{\small{II} H and K lines are well known as indicators of stellar activity. These lines are
collisional controlled  and respond to an increase in the temperature of the lower 
chromosphere of quiet stars like the Sun or Arcturus \citep{ayres75c}. 
Hence, the appearance of these lines in the spectrum of \pr{} is the evidence 
 of chromosphere. Moreover, these lines are extremely strong:

\begin{itemize}
\item Emission lines of H and K Ca{\small{II}} fill the broad absorption lines seen in less
 active M dwarfs of the same spectral type, such as GJ699 (Barnard star) as shown in Fig.\ \ref{_ca12}. 
\item These emission lines do not show any wavelength shift with respect to the photospheric lines. 
\item The emission profiles of the H and K lines are more intense in the quiet states, in contrary to what we 
see for Balmer lines of hydrogen. Likely, big flares
might affect the regions of their formation.
\item The self-absorptions seen in the cores of the Ca{\small{II}} H and K lines indicates
that the temperature drops at the upper boundary of their formation region.
\item The intensity of their components changes in time and with activity, as demonstrated
in the movie\footnote{ftp://ftp.mao.kiev.ua/pub/yp/2017/p/cahk.avi} in the supplementary material.
\end{itemize}

\begin{figure*}
\centering
\includegraphics[width=75mm, angle=0]{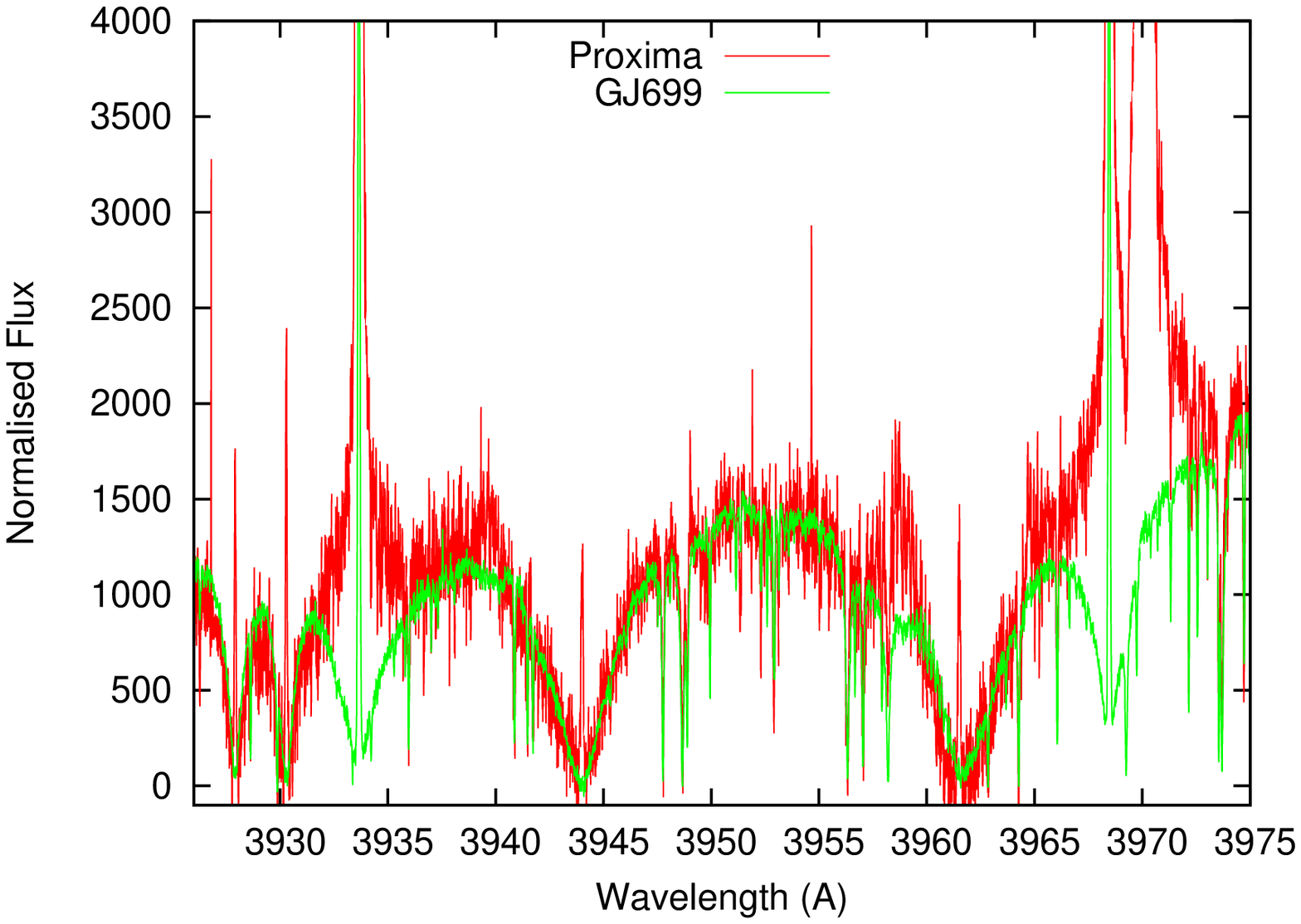}
\includegraphics[width=75mm, angle=0]{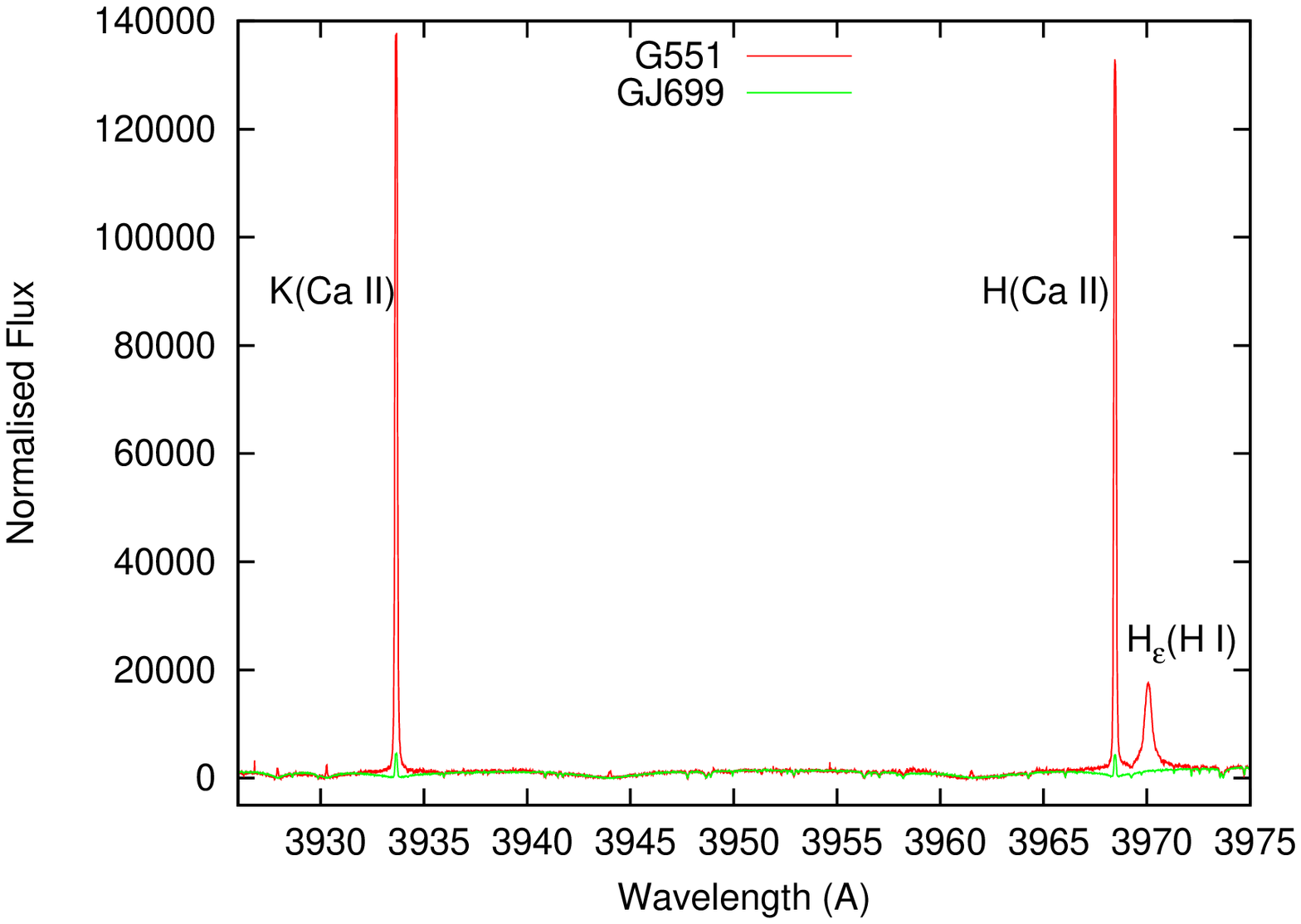}
\caption{{\it Left}: emission lines in the spectral region of the Ca II H and K  lines in of  
the star Proxima and and GJ\,699\@,  see section \protect\ref{_caii}.
{\it Right}: the same plot shown over a wider wavelength range. The 
GJ699 spectrum shown here was downloaded from the HARPS archive.}
\label{_ca12}
\end{figure*}

\subsubsection{Na H and K resonance lines}
\label{_caii}
Strong emission H and K resonance lines of sodium are notable features
in the observed spectra of \pr{}. These lines are controlled by photoelectric processes 
(Thomas 1959). For these lines,  the ratio of the photo-ionisation sink to the collisional sink 
is much larger than 1, see \cite{athay72}.
Therefore, these lines, like the hydrogen lines, show a rather marginal response to the temperature gradients 
present in chromospheres. We refer the reader to the response of lithium line on 
chromospheric-like structures in the atmosphere of M dwarfs described in \citet{pavlenko98a} . 
Only extreme cases of chromospheric activity can provide emission cores in the 
photo-electrically controlled lines, which is most likely the case of \pr{}.

In Fig.\ \ref{_na12}, we compared the profiles of K and H lines of the Na{\small{I}} resonance doublet, 
which indicate a few non-trivial results: 

\begin{itemize}
\item The strong emission cores seen in the Na{\small{I}} resonance doublet show significant
changes of total emitted energy, see 
movie\footnote{ftp://ftp.mao.kiev.ua/pub/yp/2017/p/na5890.avi}. 
\item However dispersion of velocities in line forming region change very little with the activity level. 
In the more quiet mode (QC), emission profiles of both components are a bit narrower,
while flares increase the width of the emitted lines. Likely, resonance doublet of Na{\small{I}} 
forms in the chromosphere of the star.
\item The profiles of strong photospheric absorption lines practically do not show any response 
on the level of activity, suggesting that photospheric layers are not bound or weakly
bound with regions governed by stellar activity processes.
\end{itemize}

\begin{figure*}
\centering
\includegraphics[width=75mm, angle=0]{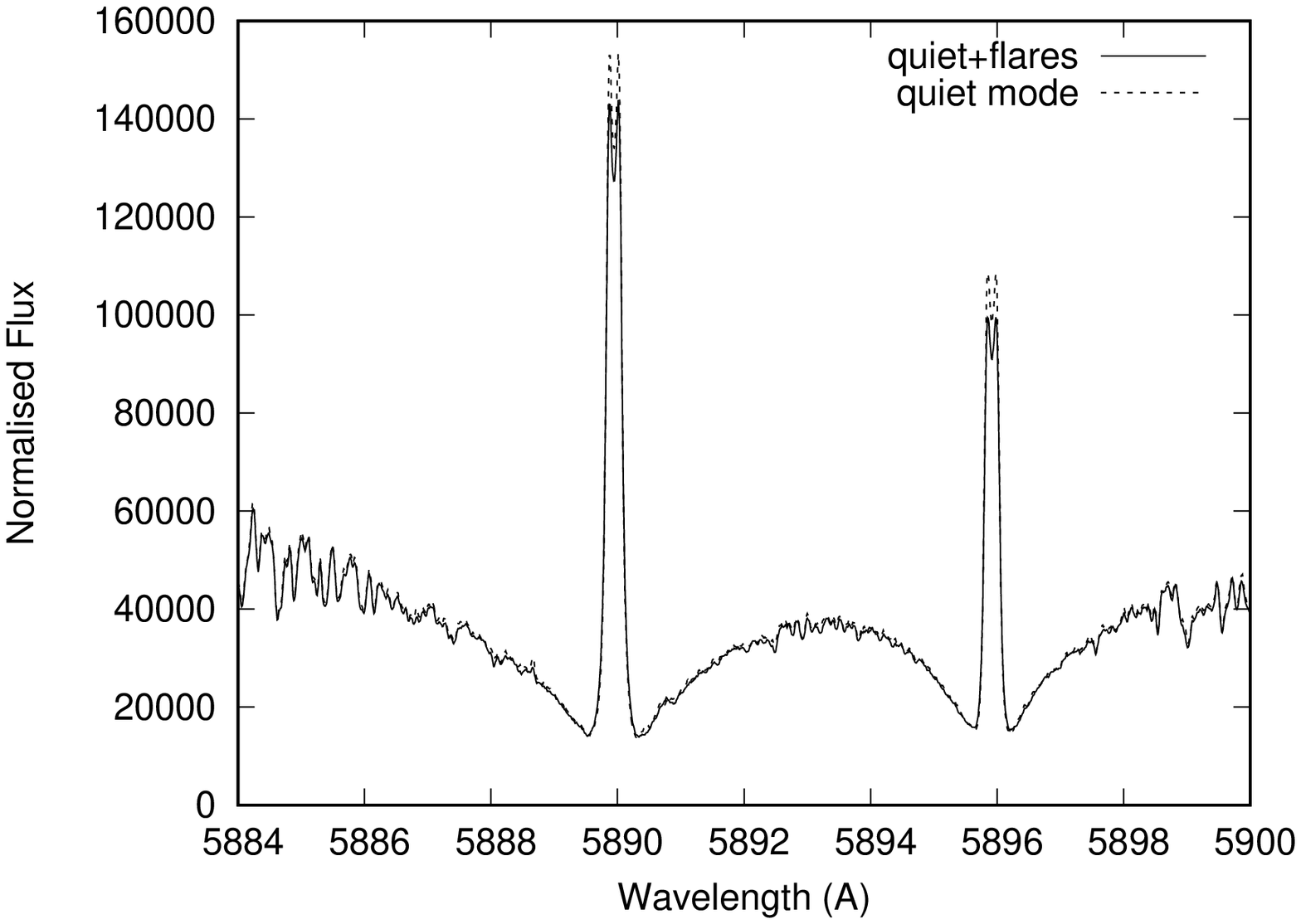}
\includegraphics[width=75mm, angle=0]{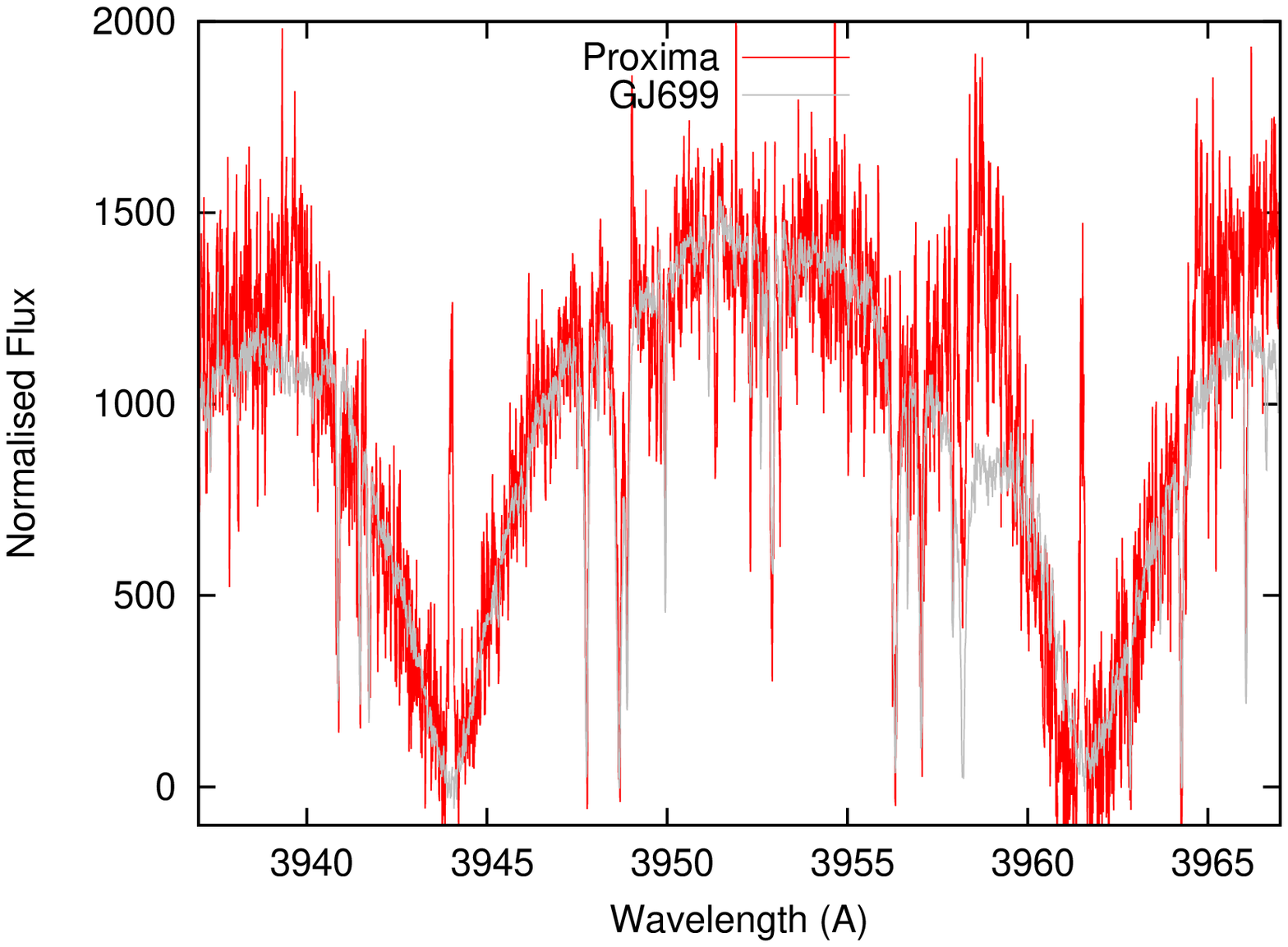}
\caption{{\it Left}: resonance doublet of Na{\small{I}} in different levels of activity.
{\it Right}: resonance doublet of Al{\small{I}} in comparison with GJ\,699, both shown in 
quiet state.}
\label{_na12}
\end{figure*}

\subsubsection{Other emission lines}
\label{_otherl}

In Table \ref{_other} we provide the list of other emission lines with emission cores seen in 
the spectrum of Proxima\@. We find emission lines of many different elements, from  
He{\small{I} (z=2) to Dy (z=66). Our list is more complete than the one given in \citet{fuhr11} 
because of the higher spectral resolution of our spectra. Moreover, our list contains both
pure chromospheric lines and lines of very high excitation like He{\small{I}}, which can be 
formed in areas governed by shock waves (i.e.\ formed outside the  chromosphere). 
We can distinguish them by the widths of their observed profiles because chromospheric 
lines formed in the cores of absorption lines are narrower than the lines of hydrogen and 
helium formed in the layers with larger velocity dispersion.
    
We assign some specific labels to some of the lines, as follows:
-- 'ec' -- absorption lines with emission cores,

-- 'ecs' -- emission cores with self-absorption as shown in the left panel of Fig.\ \ref{_56},

-- P Cyg and iP Cyg - absorption lines with emission components in the red and blue wings, 
respectively. An inverse P Cyg line is shown in the right panel of Fig.\ \ref{_56}. 

We generally find that most of the strong resonance lines show emission cores and 
are often shifted with respect to the central wavelengths. Observed emission 
cores do not form in the spherically symmetrical and stable atmosphere. Moreover, 
the strong temporal changes of the emission lines provide evidence that the 
layers of the chromosphere where they form are strongly affected by 
the flaring processes. Some other cases show P Cyg or inverse P Cyg profiles. 
These phenomena likely reflect the complicate dynamical processes occurring
in the atmosphere of Proxima\@.

\begin{figure*}
\centering
\includegraphics[width=75mm, angle=0]{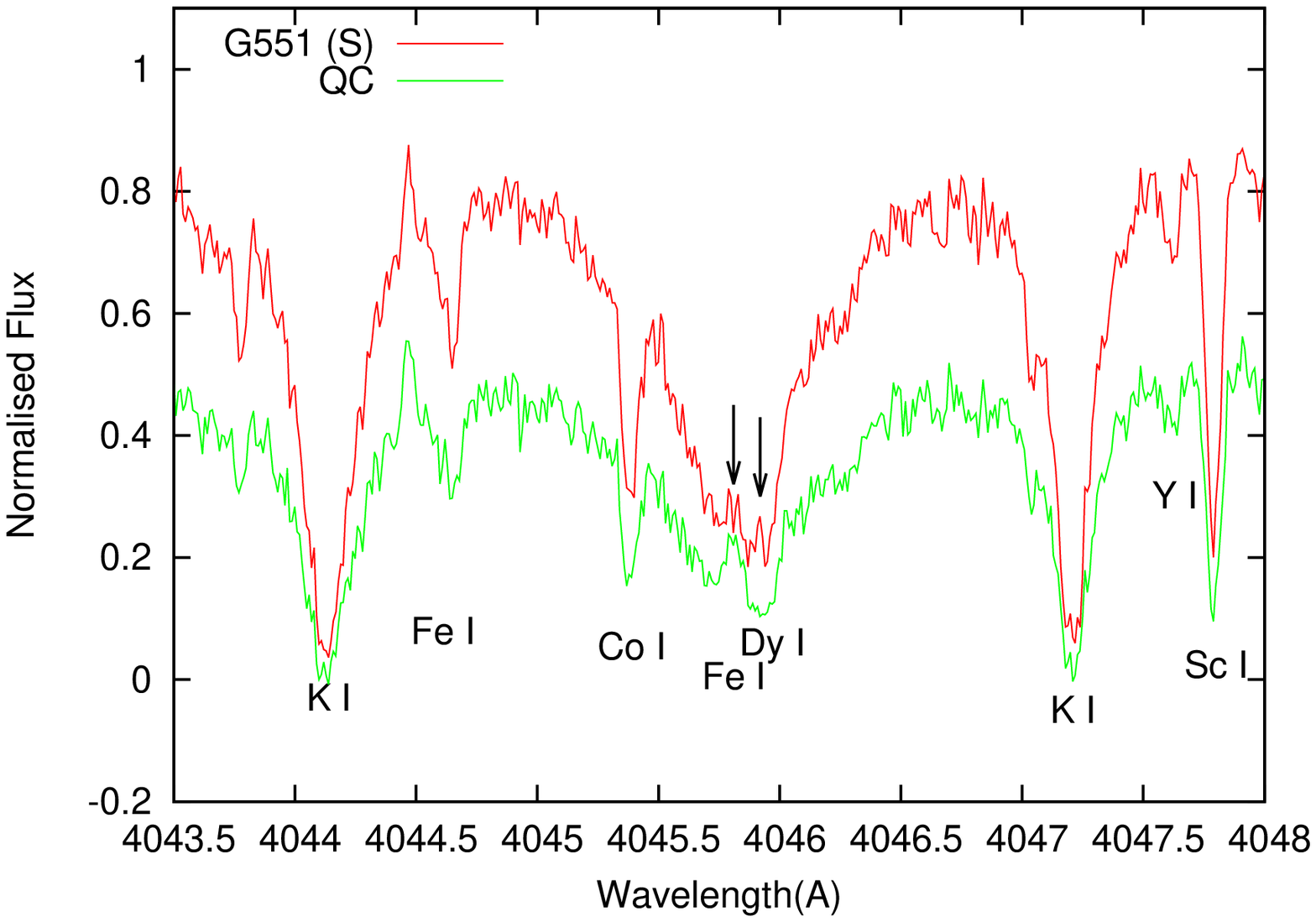}
\includegraphics[width=75mm, angle=0]{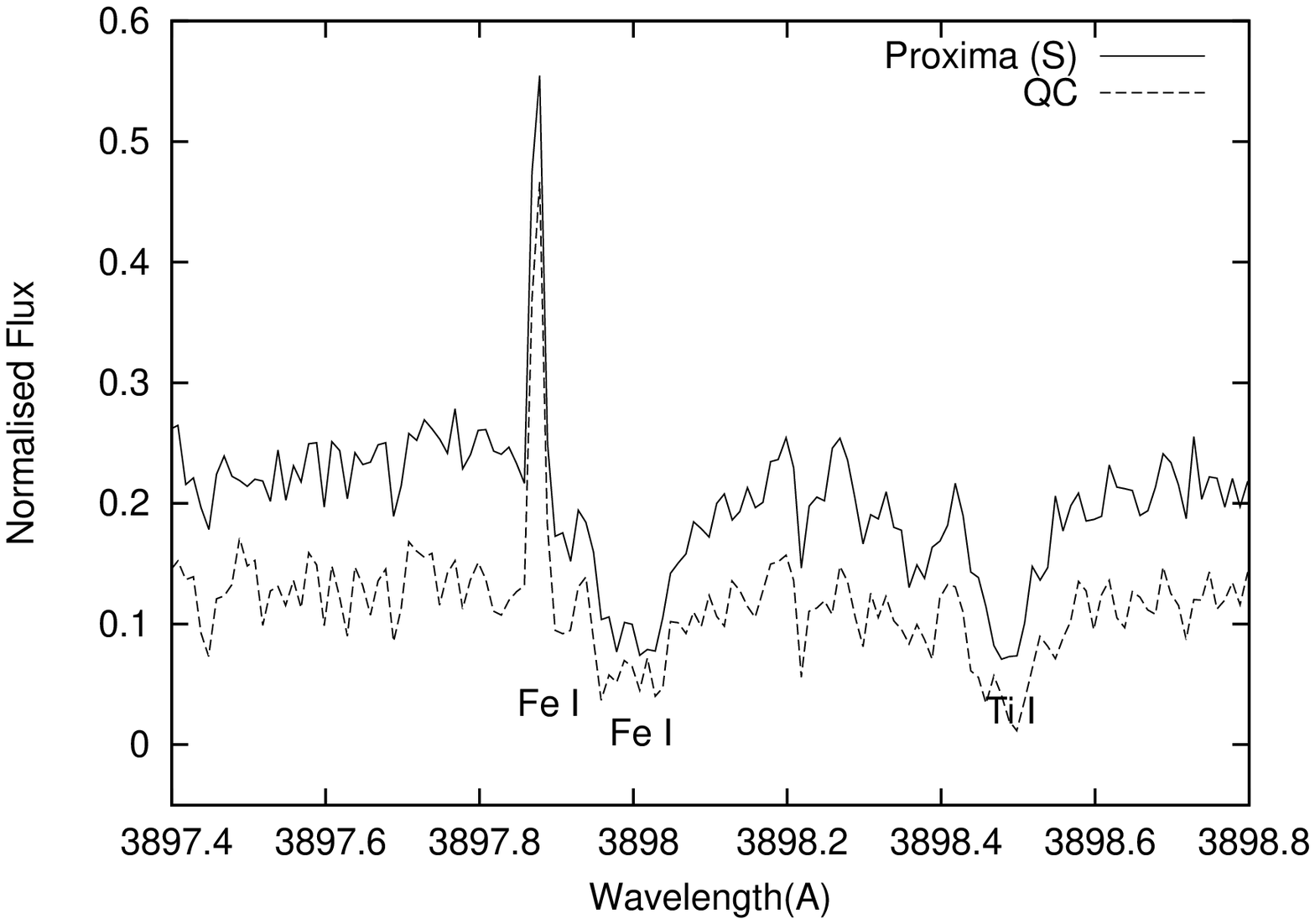}
\caption{{\it Left}: absorption lines of the neutral metals with 
emission cores. 
Left arrow at 4045.81\AA{} marks the self-absorption in the core of the strong Fe I 
absorption line. In the 'S' spectrum we note the strong emission detail 
in the core 
of the nearby Dy I $\lambda$ 4045.97\AA{} line is marked by the right arrow. 
In the 'QC' spectrum the detail 
is ''eaten'' by self-absorption. Self-absorption in the weak emission core of K I line at
$\lambda$ 4047.21 is seen as well.
{\it Right}: example of iP Cyg profiles. }
\label{_56}
\end{figure*}

%
\section{Discussion on the atmosphere of \pr{}}
\label{ProxCen:discussion}

\subsection{Photosphere}

Analysis of the VLT/X-shooter spectra obtained with intermediate resolution showed
that the atmosphere of Proxima in the optical and near IR spectral range is similar to the rest of normal
M-dwarfs of the same spectral classes M5-M6. The optical
spectral range is governed by Ti and VO bands, molecular bands of water dominates in the near IR.  
Our theoretical spectra computed for a model atmosphere of \Tef{} = 2900 K fits well enough the 
observed spectral energy distribution in accordance with 
 results of \cite{rajp11} and \cite{pass16}. 
In summary, we obtain rather good fits of our synthetic spectra computed for the canonical
PHOENIX model atmosphere of solar metallicity across all spectral ranges observed by \vlt{}, except 
for slightly lower fluxes in the spectral region of the K-band,
which seems to be a common problem of similar mid M-dwarf investigations, see 
\cite{pavl06}. 

Our computed  spectral energy distributions depend on log g rather marginally, still we note 
a weak trend toward lower log g in accordance with \logg = 4.5 found by \cite{mann15}. 
On the other hand, strong alkali lines provide clear response 
on log g changes. We obtained
better fits to profiles of some observed absorption lines of Na, Rb, subordinate
triplet of Na at 8200\AA{} with log g = 5.0, see also \cite{rajp11}, 
instead of log g = 5.5, as obtained by \cite{pass16}. Our result is in agreement 
with the log g = 5.23\,$\pm$\, 0.14 derived from the interferometric measurement 
of the Proxima radius and  mass-radius relations \citep{demo09}.

The opacity of TiO bands around 4200\AA{} starts to decrease bluewards \citep{pavl14}. 
At shorter wavelengths, we explore deeper regions of the atmosphere of Proxima, where
the observed spectrum is dominated by absorption lines of neutral atoms. The resonance lines
of neutral species become very strong. Theoretical spectra provide too strong wings due to 
the effects of pressure damping which broadens these lines. In the case of \pr{} we obtain  
the best solution for $\kappa$ = 80. This value looks very high, but we work in the regime 
of low temperatures where the conventional opacity is extremely low in this case due to the low density 
of the $H^-$ ion, which is the main source of opacity in the atmospheres of M stars. In other words,
we should find larger electron densities to increase opacity in the low photospheric layers. 
We suggest that this effect may be created by the over-ionisation of alkali metals which are 
the main donors of free electrons. Indeed, photons with $\lambda <$ 4200\AA{} can ionise
the neutral atoms in late-type atmosphere due to the low electron densities where processes 
of recombination are not so effective. 
The additional opacity affects only the emitted spectrum in the UV and blue wavelengths,
occurring deep in the atmosphere. The temperature structure of the upper layers of the 
model atmosphere are mainly determined by molecular opacities, which explain why we 
can reliably fit the spectral energy distribution of \pr{} at longer wavelengths. 

 On the other hand, for some spectral diagnostics, it could also be that the
analysis is too simplified and 1-D LTE syntheses may indeed
underestimate some of the opacity (lines) in the blue. More
complex analysis involving 3-D atmospheres from hydro simulations
 sometimes find that the missing opacity is even larger
 in 3D NLTE models than in 1-D LTE. There are hints in the paper
by \cite{fuhr11} that much more sophisticated 3-D atmospheres should be
 applied in the case of Proxima.

\subsection{Chromosphere} 

The chromosphere represents the extended part of the atmosphere of Proxima, where narrow 
emission lines of different intensities vary with time. We observe temporal variations of these 
lines because of interactions between the flares and the chromosphere. We know that solar
flares originate in high chromosphere-corona regions. The strongest flares move downwards
to get deeper into the photosphere. However, our movies show mainly temporal changes 
of intensity of chromospheric lines, while the photospheric spectrum shows marginal 
response to flares. In the most cases, even the strongest flares seem to occur in local 
regions far above the surface/photosphere of Proxima\@. The TiO line forming region is 
separated from the hot flare regions by the mantle of the cool plasma. Nevertheless, 
the chromosphere of Proxima is very powerful because it shows many absorption lines 
of neutral metals with narrow emission cores. 
These emission cores most likely form in the 
chromosphere, above the temperature minimum. 
The presence of emission cores within the lines of neutral alkalies and neutral atoms which 
are fully controlled by radiation can only occur in the presence of steep temperature gradients
in the chromosphere, see e.g.\ NLTE simulations of emission cores of Li in \citet{pavl98}.
 
\subsection{Flares} 

Hydrogen Balmer lines in emission is a common phenomenon in the Sun. 
Balmer lines are controlled by photo-electrical processes and are not sensitive to the temperature structure of 
the atmosphere. Their emission is mainly the result of ionization and collisional processes 
created by shock waves following flare event. The Hydrogen emission lines seen in the Proxima
spectrum form mainly in flare regions. They are broader than chromospheric lines 
of Ca{\small{II}},  Na{\small{I}} and other neutral metals. This effect most likely results
from a larger scatter in their velocity distribution, as seen in Fig.\ \ref{_ca12}, where we 
compare the Ca{\small{II}} and H$_{\epsilon}$ lines: the Ca{\small{II}} is much stronger, but 
H$_{\epsilon}$ is broader.  

\subsection{Stellar wind} 

We observe a quasi-stationary component in the blue wing of the \Ha{} and $H_{\beta}$ emission in the HARPS
spectra. 
We interpret its presence as a result of the flow of highly ionized plasma with a
velocity of $V_{r}$\,=\,$-$30 \kmps. 
We assume that the observed component most likely 
relates to the the hot stellar wind  outflow generated by the high level of flare activity in Proxima. 
From our estimate of the total energy emitted in the $H_{\alpha}$ line, we infer a lower limit 
of the mass loss of $\dot{M} = 1.8 \times 10^{-14} M_{\odot}/yr$ because we do not consider 
emitted energies by other emission lines. 

\citet{wood00, wood01} estimated a four or six times lower mass loss of $\dot{M} \leq 0.2 \dot{M}_{\odot}$
 = 0.5 10$^{-14} M_{\odot}/yr$ or 0.3 10$^{-14} M_{\odot}/yr$ for the cases of average solar wind mass 
loss $\dot{M_{\odot}}$ = 1.374 10$^{12}$ g/sec = 2.3~$10^{-14} M_{\odot}/yr$ \citep{hund97} and
$\dot{M_{\odot}} \sim  1.3~10^{-14} M_{\odot}/yr$ \citep{gold96}, respectively. Nevertheless, 
measurements by \cite{wood00, wood01} are based on the analysis of the $L_{\alpha}$ 
absorption which relates  to absorption by neutral hydrogen atoms. Both estimates relate to different 
parts of the stellar wind. 
It is worth noting that the level of activity and, respectively, mass loss 
for the late type stars should change in time.
Indeed, we observe different levels of activity in the M-dwarfs population of our Galaxy.  The 
spectrum of GJ699 used in our paper for the comparison with Proxima provides 
clear evidences of much lower level of activity.
Likely, nowadays \pr{} passes its evolutionary epoch of high activity. 

In our spectra we see manifestations of cool and hot components of the stellar wind from \pr{}.
Cool neutral hydrogen located above the flare region provides the 
asymmetrical self-absorption in \Ha{} core, as discussed in \citet{fuhr11}.
We defer a more detailed analysis of this phenomenon 
to a future paper.

\section{Conclusions}

In the framework our work at least a few results were obtained: \\
$\bullet$ From the fits spectral energy distributions observed in the optical and near
infrared spectral ranges we obtained effective temperature of \pr{} ~ \Tef{} = 2900\,$\pm$\, 100 K. \\
$\bullet$ Fit to profiles of strong atomic lines atomic lines observed in the optical
spectrum of \pr{} provides good restriction for the gravity in atmosphere 
\logg = 5.0\,$\pm$\, 0.25. \\
$\bullet$ From the analysis of strong resonance and subordinate lines of Na, K, Rb as
well as lines of intermediate strengths of Ti I and Fe I lines formed
at the background of TiO and VO bands we obtain 
 solar abundances of these elements in the atmosphere of \pr{}.  \\
$\bullet$ From the fits to the observed spectrum across Li resonance doublet we determined the
upper limit of Li abundance log N(Li) = $-$12.04, which is consistent with the expected depletion
 of an old fully convective low-mass star.  \\
$\bullet$ Photospheric lines observed in the optical and infrared spectra of Proxima can be fitted 
by standard synthetic models. However observed strong lines of low excitation potentials
in the blue spectral region show narrower profiles than expected indicating  a formation 
in lower pressure layers in the atmosphere.
 We were able to reproduce their profiles incorporating
additional opacity, which shifts their formation layers upwards, in the lower pressure regions. \\
$\bullet$ In spite of a comparatively high level of activity 
we found that the photospheric spectrum show rather marginal response on the flare
activity of Proxima, except for very strong flares. \\
$\bullet$ The emission lines of hydrogen are good indicators of stellar activity.
At the times of strong flares they become more intense. 
On the contrary, strong emission lines formed in the chromosphere, i.e., 
H\&K Ca II, H\&K Na I, reduce their intensity at the occurrence of strong flares.
Likely strong flares change the structure of chromosphere. These chromospheric originated
lines are narrower in comparison with hydrogen lines formed in the flare region.\\
$\bullet$ The He I line at 4026.19\AA{} is observed in emission and reduces its intensity 
in the presence of strong flares,too.
Due to the larger excitation potential of the He I line, it should form
in hotter layers above the hydrogen lines formation region. 
In the absence of strong flares the He I emission line at 4026.19\AA{} shows multicomponent profile.
Likely, it reflects complicate structure of the line forming region. 
In the 'S' spectrum the intensity of the He I lines is lower than in QC,
it means that the He I lines formation layers are affected by flares. \\
$\bullet$ In the blue wing of emission hydrogen \Ha{} and $H_{\beta}$ lines we found an emission component
shifted to \Vr = 30 \kmps{} with respect to their cores. 
We interpret the emission components as evidence of the hot stellar wind from  \pr{}.
Using the simple model of complete ionisation of hydrogen atoms in the stellar wind we
estimate a minimum mass loss of $\dot M$ = 1.8 10$^{-14} M\odot/yr$. \\
 
\begin{acknowledgements}
Based on observations collected at the European Organisation for Astronomical Research 
in the Southern Hemisphere under ESO programme(s) 087.D-0300(A).
This is research has made use of the services of the ESO Science Archive Facility.

YP thanks financial support from the Fundaci\'on Jes\'us Serra for a 2 month stay 
(Sept--Oct 2016) as a visiting professor at the Instituto de Astrof\'isica de Canarias (IAC) 
in Tenerife. NL and VJSB are supported by the AYA2015-69350-C3-2-P program from 
Spanish Ministry of Economy and Competitiveness (MINECO). 
J.I.G.H. acknowledges financial support from the Spanish MINECO under the 2013 
Ram\'on y Cajal program MINECO RYC-2013-14875, and A.S.M., J.I.G.H., and R.R.L. 
also acknowledge financial support from the Spanish ministry project MINECO AYA2014-56359-P. 

The authors kindly thank M.A. Bautista, S.N. Nahar, M.J. Seaton, D.A. Verner who supplied 
data compiled in the NIST database.
This research has made use of the Simbad and Vizier databases, operated
at the Centre de Donn\'ees Astronomiques de Strasbourg (CDS), and
of NASA's Astrophysics Data System Bibliographic Services (ADS).

We thank the anonymous referee for his/her thorough
review and highly appreciate the comments and suggestions, which
significantly contributed to improving the quality of the publication.

\end{acknowledgements}

%
%
\bibliographystyle{aa}
\bibliography{mnemonic,biblio_ProximaCen}

\begin{appendix}
\section{}
\begin{figure*}
  \centering
  \includegraphics[width=0.48\linewidth, angle=0]{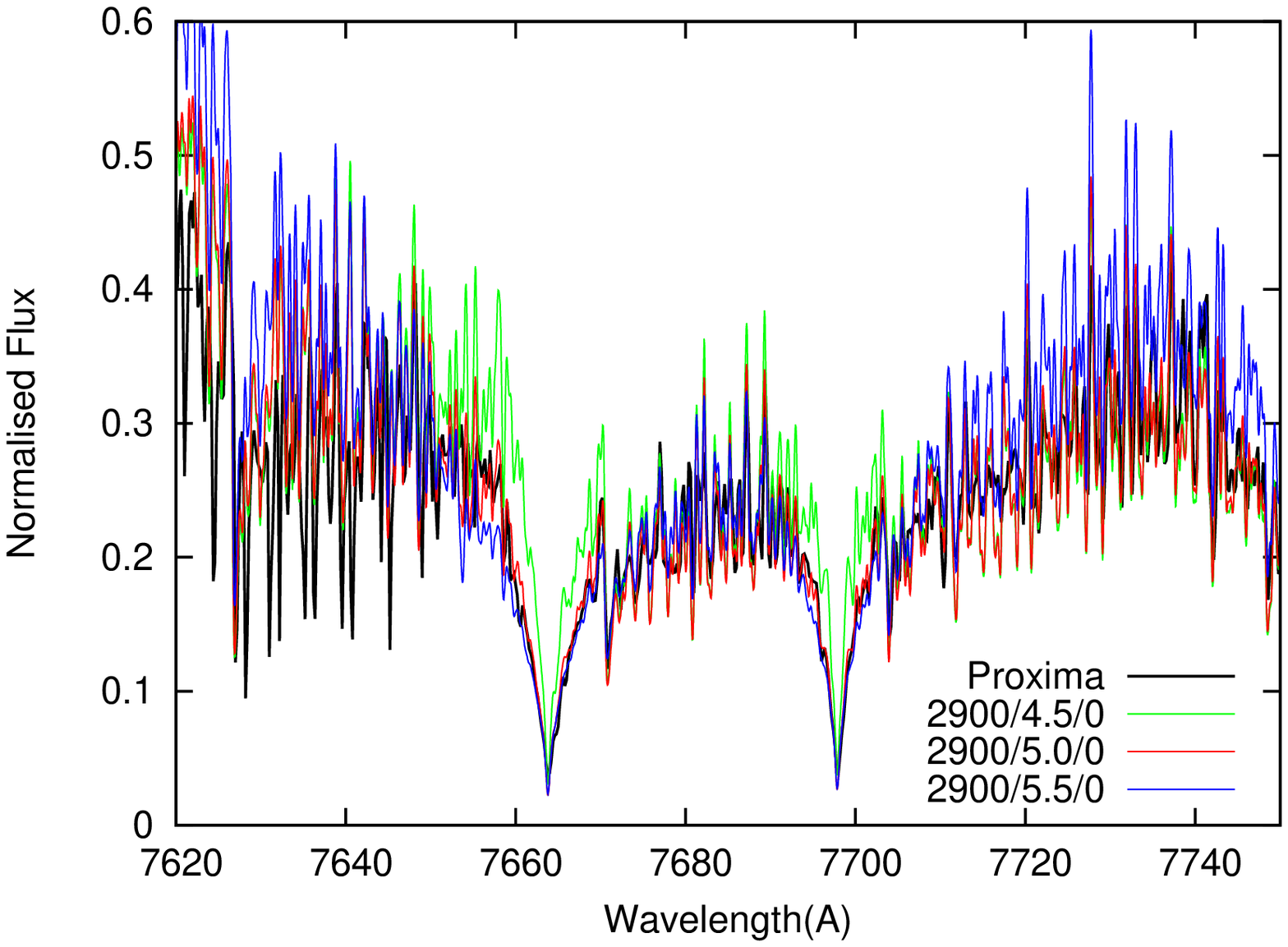}
  \includegraphics[width=0.48\linewidth, angle=0]{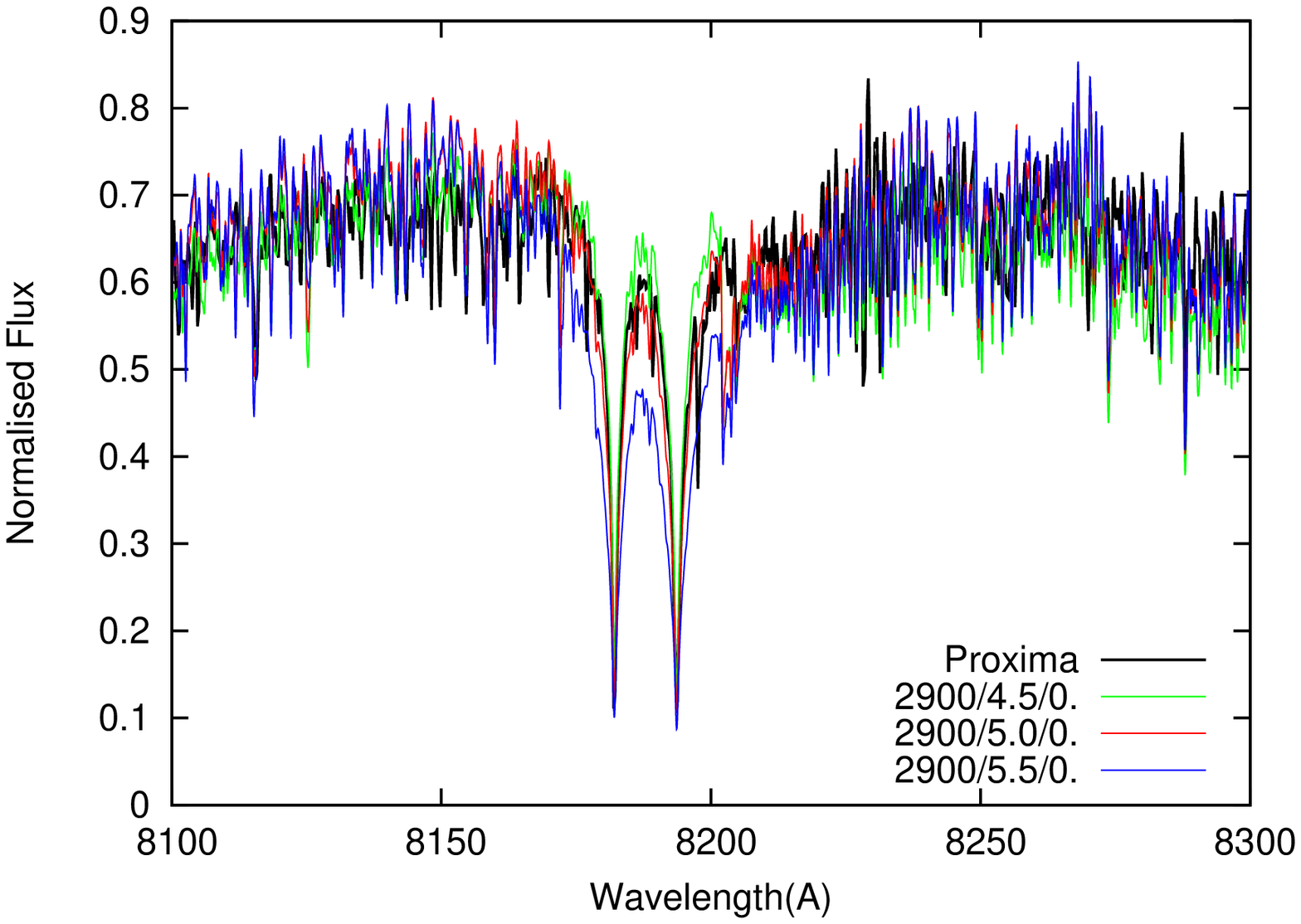}
  \includegraphics[width=0.48\linewidth, angle=0]{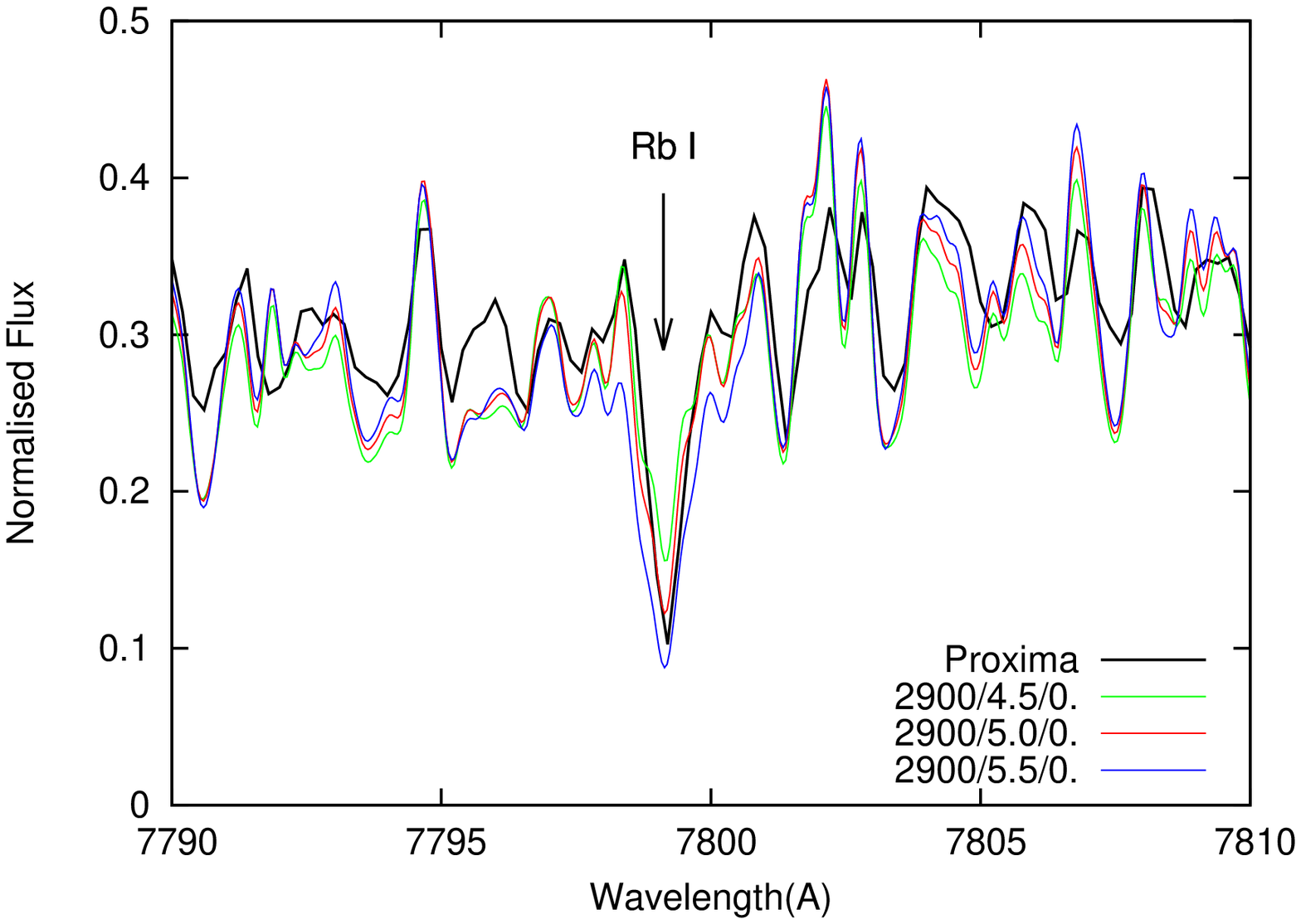}
  \includegraphics[width=0.48\linewidth, angle=0]{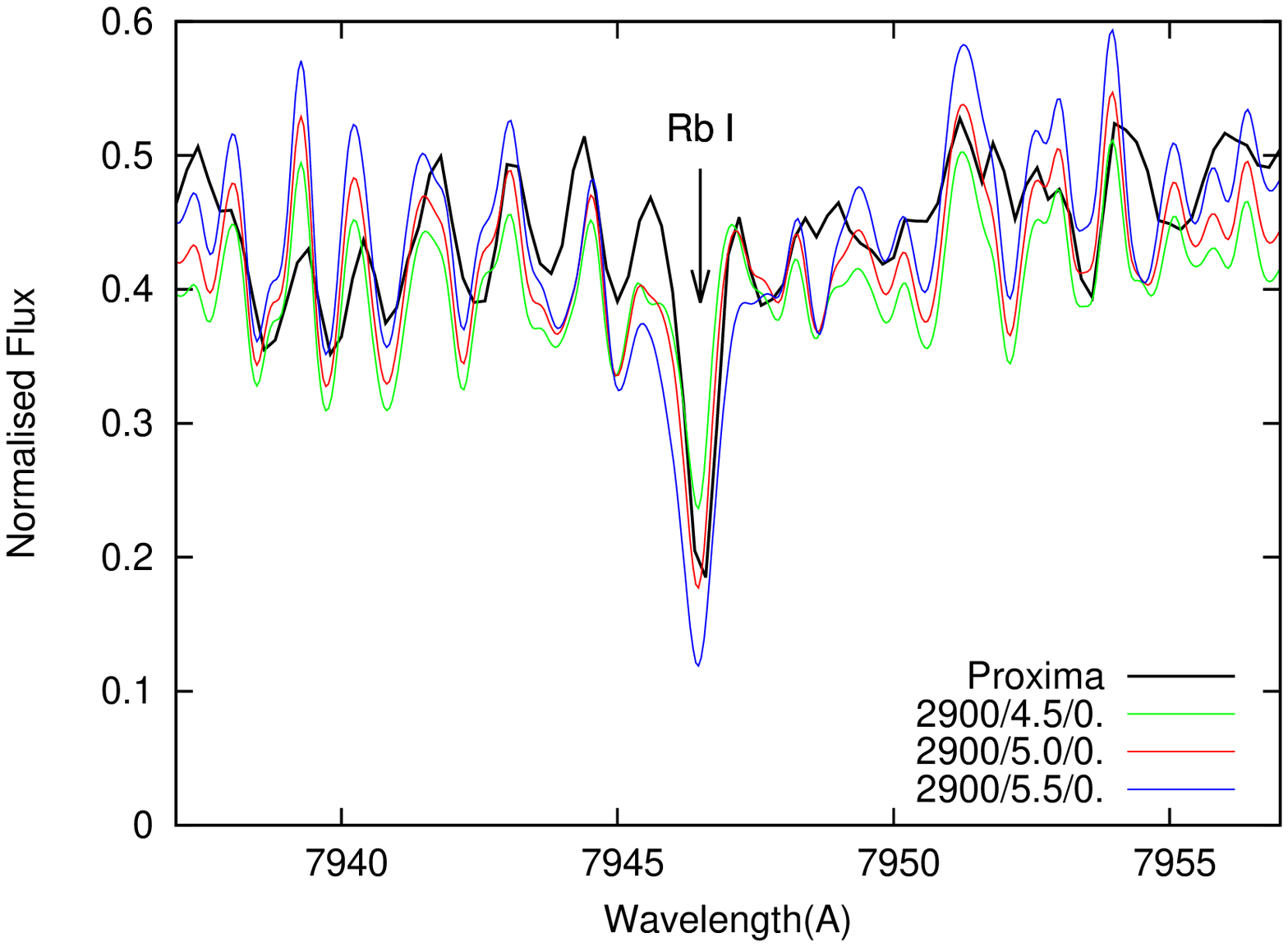}
  \includegraphics[width=0.48\linewidth, angle=0]{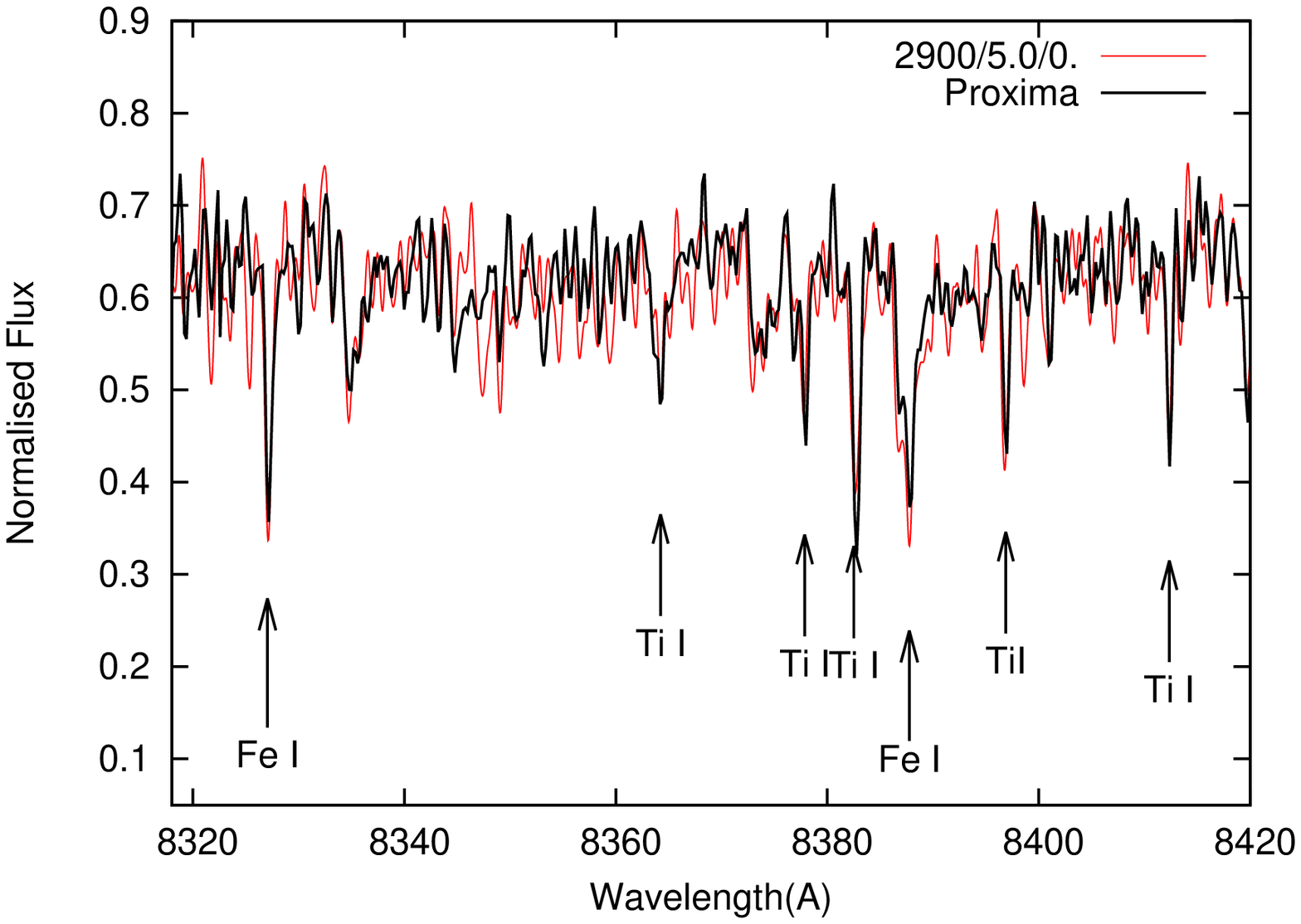}
  \includegraphics[width=0.48\linewidth, angle=0]{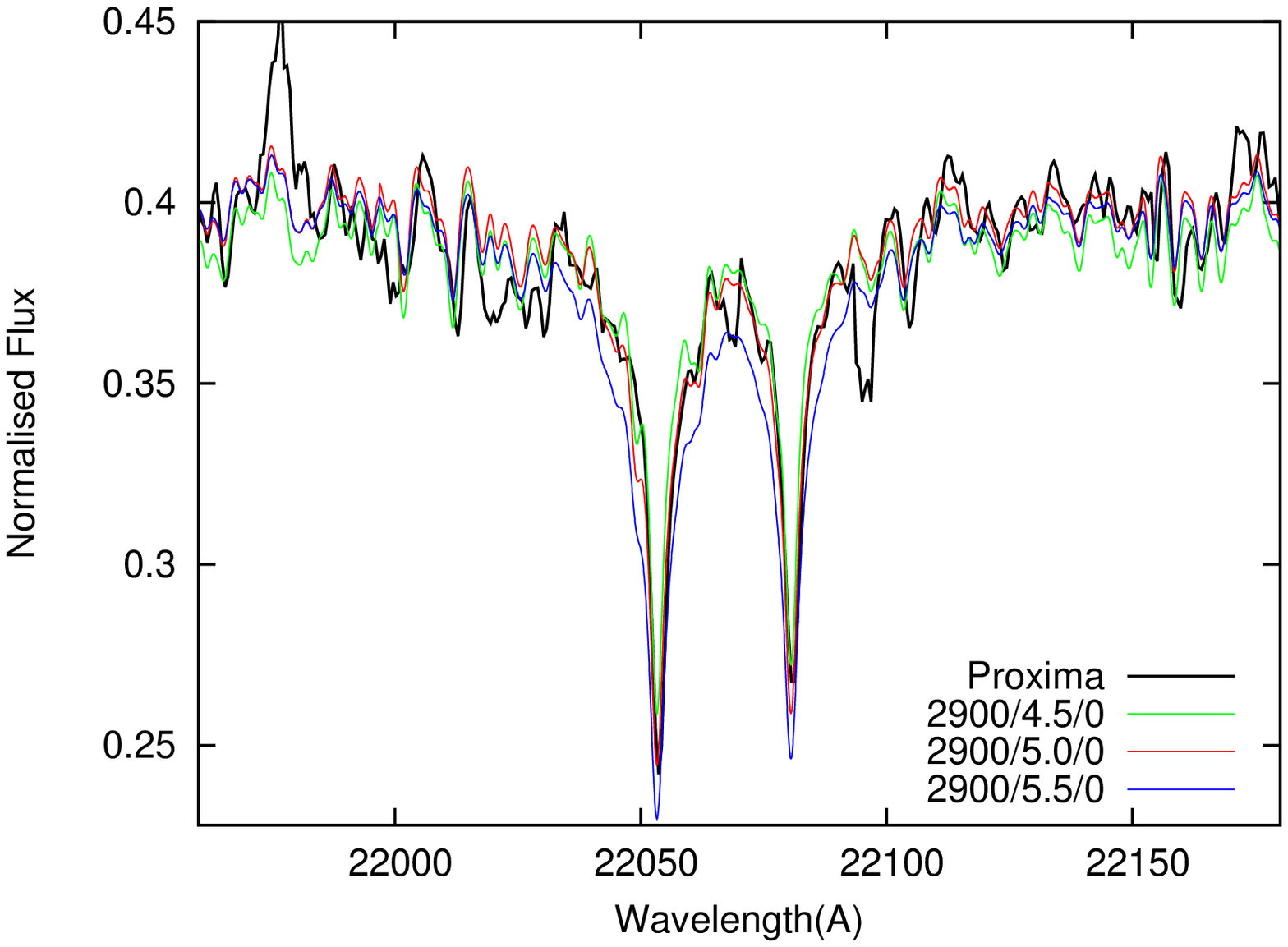}
   \caption{{\it Top panel}: fits to the observed K{\small{I}} 
resonance doublet (left), 
   subordinate triplet of Na{\small{I}} (right).
{it Middle panel}: fits to Rb{\small{I}} resonance doublet lines.
 {\it Bottom panel}: fits to the observed profiles of
   Ti{\small{I}}, Fe{\small{I}}, and Na{\small{I}} lines.}
   \label{_atop}
\end{figure*}

\begin{figure*}
  \centering
  \includegraphics[width=0.9\linewidth, angle=0]{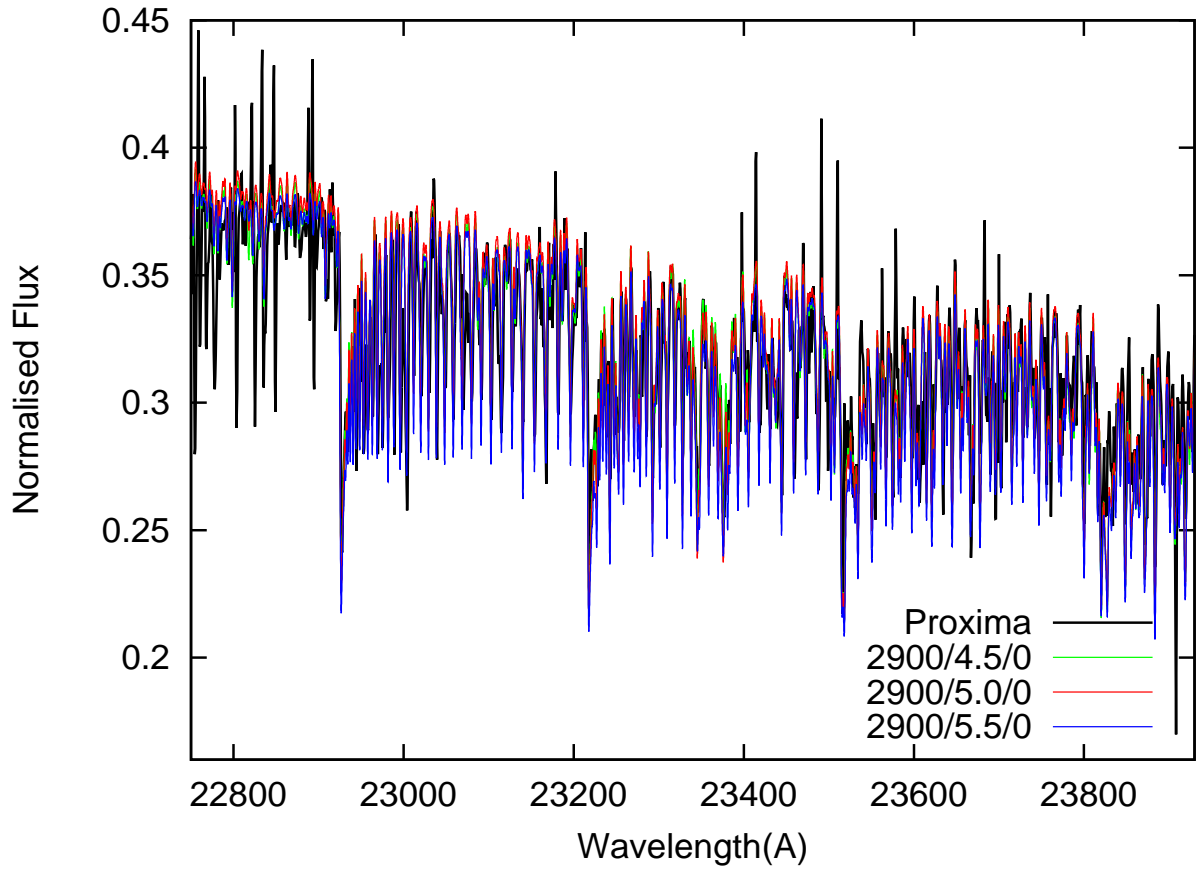}
    \includegraphics[width=0.9\linewidth, angle=0]{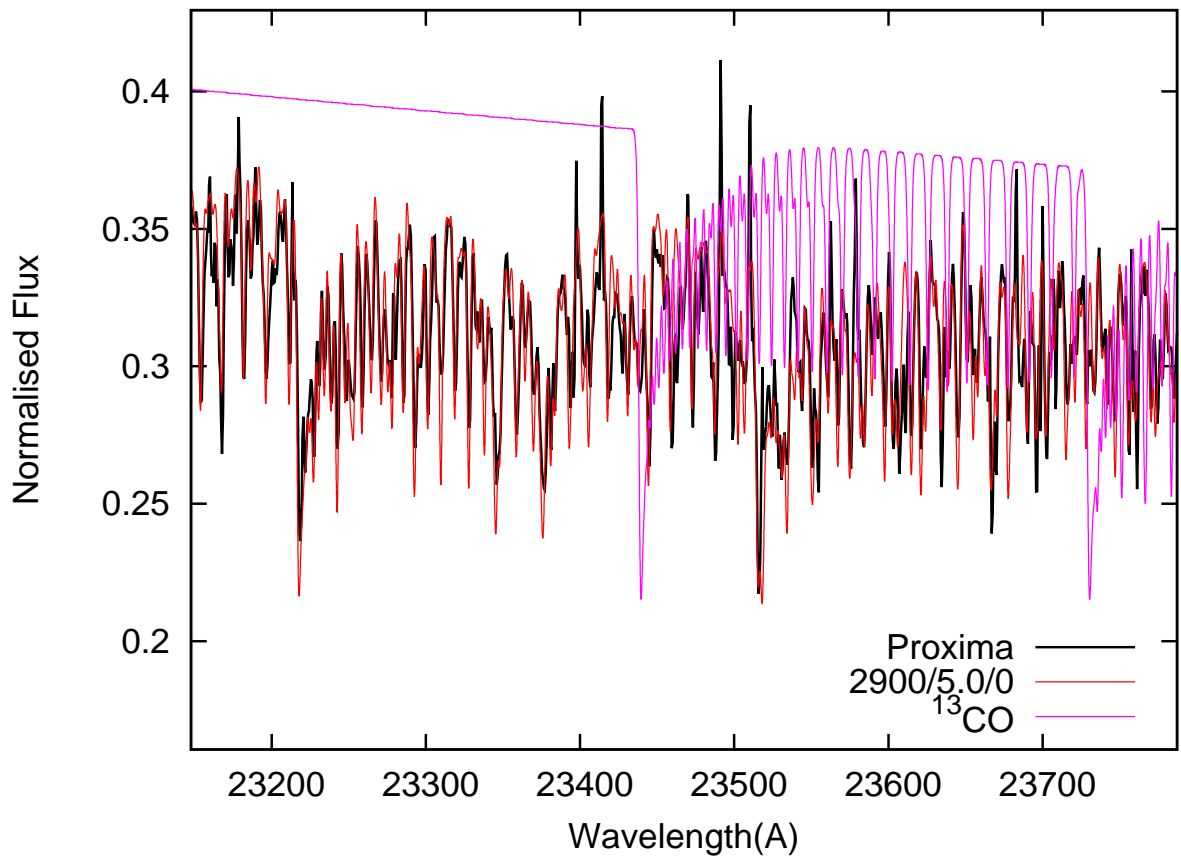}
   \caption{{\it Top}: fit of our synthetic spectra governed mainly 
by $^{12}$CO bands to the observed \vlt{} spectrum. {\it Bottom}: $^{13}$CO bands 
in the theoretical spectrum computed
   for the parameters of Proxima\@.}
   \label{_CO}
\end{figure*}

\begin{figure*}
\centering
\includegraphics[width=0.9\linewidth, angle=0]{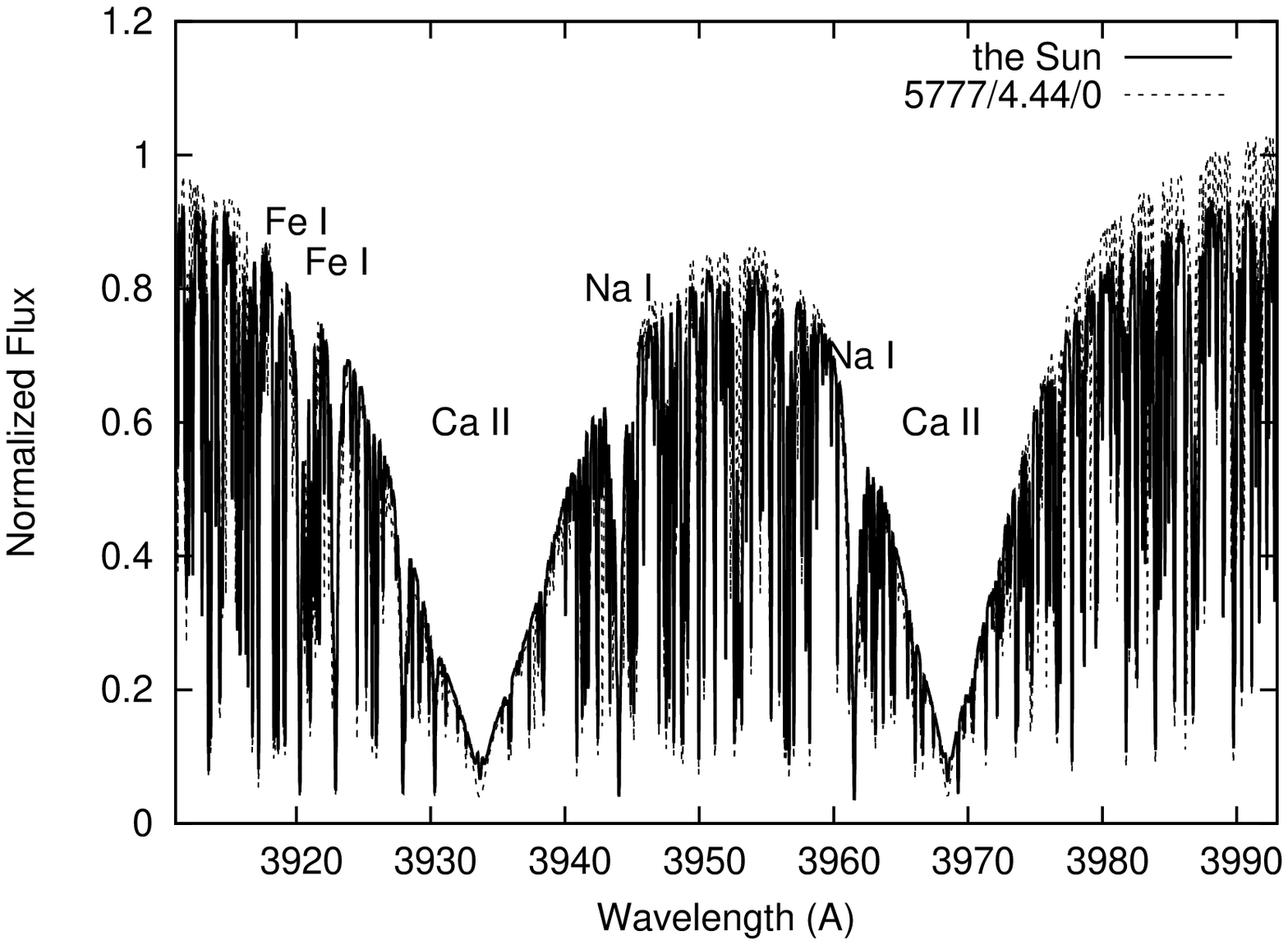}
\includegraphics[width=0.9\linewidth, angle=0]{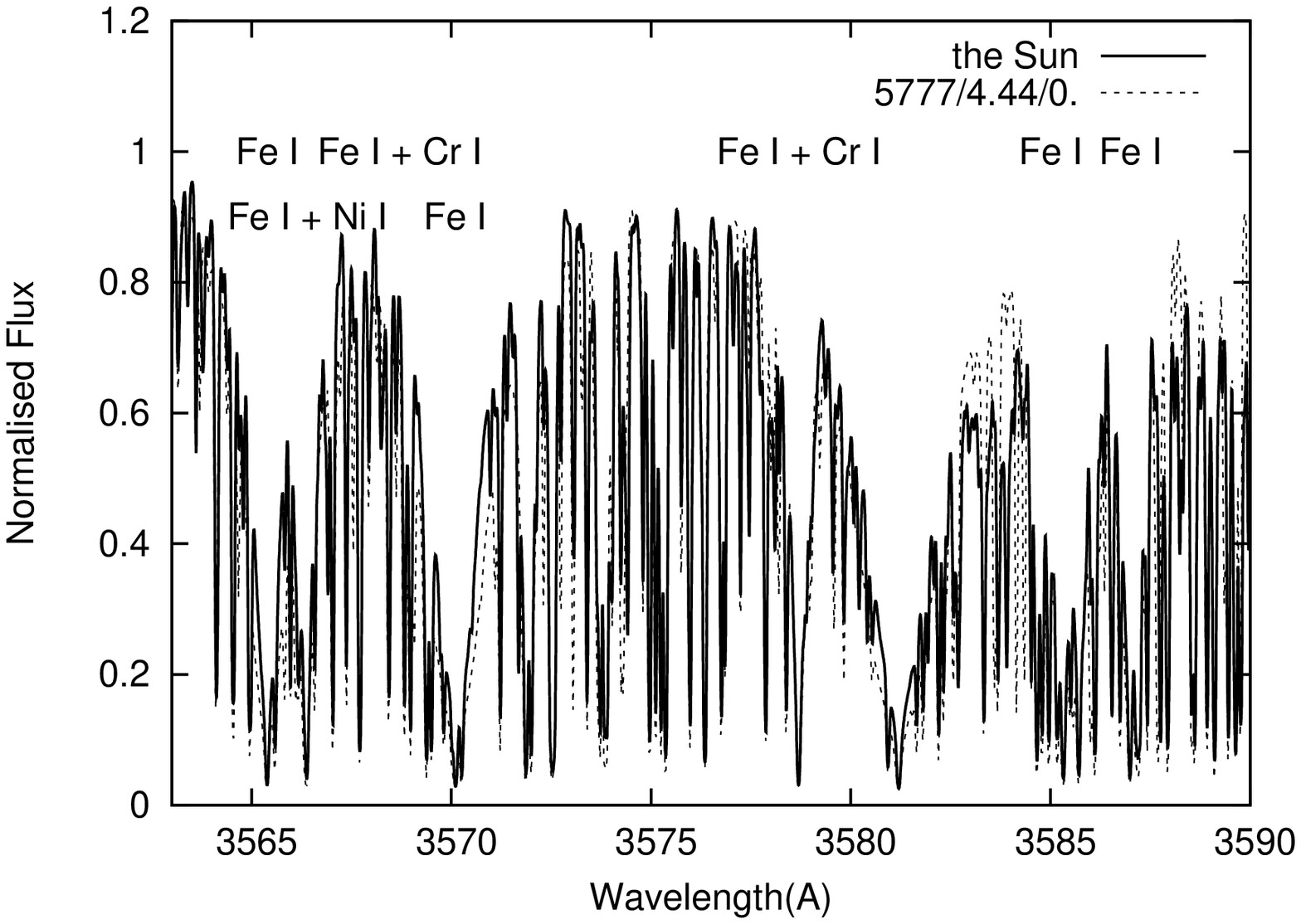}
\caption{Comparison of the strong lines in the spectrum of the Sun as a star \cite{_kuru84} 
and the theoretical spectrum computed for 1D model atmosphere of \Tef/\logg/[Fe/H] = 5777/4.44/0.0\@.}
\label{_sun}
\end{figure*}

\begin{onecolumn}

\begin{longtable}{c c c c c c} 
\hline\hline
 \caption{Emission lines in Proxima spectrum, see section\ref{_otherl}} \label{_other} \\
 \hline\hline
$\lambda$   & Element  &   $gf$      & E" (eV)&   &Remarks \\
 \hline
        &          &              &      &      &  \\
 \hline
3819.57  &    Cr I &  1.16E+00  &  2.708 & S,QC & ecs                                           \\
3824.44  &    Fe{\small{I}} &  4.35E-02  &  0.0   & S,QC & ec                                           \\ 
3829.36  &    Mg I &  5.93E-01  &  2.709 & S,QC & ec                                             \\
3832.30  &    Mg I &  1.33E+00  &  2.712 & S,QC &  ec                                         \\  
3837.60  &         &            &        &  S,QC                                                 \\ 
3838.29  &    Mg I &  2.50E+00  &  2.717 &  S,QC & ec      \\
3840.75  &    V I &  7.28E-01  &  0.040 &   S,QC & ecs                                          \\ 
3847.33 &    V I  &  8.51E-02  &  0.017 &  S, QC&  ecs, P Cyg?                                    \\ 
3853.88  &         &            &        &  S,QC &                                           \\ 
3856.37  &    Fe{\small{I}} &  5.18E-02  &  0.052 &  S,QC &  ec                                    \\
3859.91  &    Fe{\small{I}} & 1.95E-01   & 0.000  &  S,QC &   ec                                       \\
3864.10  &    Mo I & 9.77E-01   &  0.000 &  S    & ec, iP Cyg?                        \\
3870.91  &         &            &        &   S$>$QC &                                     \\
3872.50  &   Fe{\small{I}}  & 1.18E-01   & 0.990  &   S,QC   & ec                                    \\
3878.57  &    Fe{\small{I}} &  4.18E-02  &  0.087 &   S,QC &   ec                                 \\ 
3886.28  &    Fe{\small{I}} &  8.39E-02  &  0.052 &   S,QC &                                           \\ 
3887.05  &    Fe{\small{I}} &  7.18E-02  &  0.915 &   S,QC &                    \\ 
3888.66  &    He{\small{I}}  &            &        &   S,QC &            \\
3894.03  &    Cr I &  2.24E-02  &  0.961 &   S,QC &  ec                                     \\
3894.08  &    Co I &  1.26E+00  &  1.049 &   S,QC & ec                      \\ 
3894.98  &    Co I &  3.98E-02  &  0.629 &   S,QC & ec                                            \\ 
3895.66  &    Fe{\small{I}} &  2.14E-02  &  0.110 &   S,QC  &  ec                                      \\
 3897.88 &    Fe{\small{I}} & 1.84E-01   & 2.692  &   S,QC & iP Cyg                                       \\ 
3899.71  &    Fe{\small{I}} &  2.94E-02  &  0.087 &   S,QC & ecs                                           \\ 
3903.16  &    Cr I &  5.89E-03  &  0.968 &   S,QC &  ec                                         \\ 
3903.90  &    Fe{\small{I}} &    1.57E-01&  2.990 &   QC   &                             \\
3905.52  &    Si I &  9.10E-02  &  1.909 &   S,QC &                                           \\ 
3906.48  &    Fe{\small{I}} &  5.71E-03  &  0.110 &   S,QC&    ecs                                            \\ 
3907.49  &    sr I & 2.28E+00   & 0.000  &   S,QC &   ec                         \\
3907.93  &   Fe{\small{I}}  & 7.64E-02   & 2.759  &   S,QC &   ec                          \\
3908.76  &    Cr I &  8.91E-02  &  1.004 &   S    &  ec shifted to red                                       \\ 
3909.86  &    V I, Co I &  7.94E-02  &  0.069 & S,QC & complicate blend                                        \\ 
3920.26  &    Fe{\small{I}} &  1.80E-02  &  0.121 &  S,QC & ecs shifted to red                                   \\ 
3922.91  &    Fe{\small{I}} &  2.23E-02  &  0.052 & S,QC &                                           \\ 
3926.82 &          &            &        & S,QC &  esc                                  \\ 
3927.92  &    Fe{\small{I}} &  3.01E-02  &  0.110 &  S,QC &        ec                                   \\ 
3930.30  &    Fe{\small{I}} &  3.23E-02  &  0.087 &   S,QC &         ec                                  \\ 
3939.26  &          &            &        &  S &      iP Pyg                                              \\
 3940.88 &    Fe{\small{I}} & 2.51E-03   &  0.958 &  S,QC & ec shifted to blue                              \\
 3941.49 &    Cr I & 4.07E-02   & 1.030  &   S,QC & ec shifted to blue                                   \\ 
3941.73  &    Co I &  9.33E-03  &  0.432 &  QC & ecs                           \\ 
3944.01  &    Al I &  2.38E-01  &  0.000 &  QC,S & ecs                  \\ 
3948.67  &    Ti{\small{I}} &  3.98E-01  &  0.000 & S,QC & ec                                           \\ 
3960.05  &          &            &        & S &                                               \\ 
3961.52  &    Al I &  4.75E-01  &  0.014 &   QC,S &  ecs                            \\ 
3983.13  &          &           &         & QC,S &                                             \\
4001.66  &   Fe{\small{I}}   & 1.26E-02  &2.176    & QC,S & ec blue shifted                          \\
4021.87  &   Fe{\small{I}}   & 1.87E-01  & 2.759   & QC,S & ec                         \\
4026.19  &          &            &        &  QC,S & He II                                    \\ 
4030.75  &    Mn I &  3.21E-01  &  0.0 &  QC,S & ecs                                             \\ 
4032.98  &    Ga I &  2.36E-01  &  0.0&  S &                                              \\ 
4033.06  &    Mn I &  2.27E-01  &  0.0& QC,S & ecs                                              \\ 
4034.48  &    Mn I &  1.44E-01  &  0.0& QC,S & ecC                                               \\ 
4042.93  &         &            &     & QC,S &                                \\
4044.14  &    19.00 & 1.20E-02 & 0.000 &  QC,S& P Cyg            \\
4045.97  &    Dy I &  7.08E+00  &  0.0&  S &    ec                                  \\ 
4047.21  &    K I &  6.03E-03  &  0.0 & QC,S&   ecs                   \\
4052.96  &        &            &      & QC,S &                                       \\
4062.44  &    Fe{\small{I}} & 1.37E-01  & 2.845 & QC,S & ecs                                  \\
4063.59  &    Fe{\small{I}} & 1.15E+00  & 1.557& QC   & ecs                                   \\
4064.21  &   Ti{\small{I}} & 1.20E-01 & 1.053  & QC,S & ec                                      \\
4071.74  &   Fe{\small{I}} & 9.51E-01 & 1.608  & QC,S & P Cyg                                      \\
4077.36  &    Y I &  1.89E+00  &  0.0 & QC,S & iP Cyg                                             \\ 
4077.71  &    Sr II &  1.47E+00  &  0.0 & QC,S & P Cyg                                 \\ 
4103.94  &          &            &       & QC,S &                                       \\
4111.14  &          &            &        &   na?  P Cyg, ecs                             \\ 
4115.18  &    V I & 1.18E+00 & 0.287& QC,S&   ec                                         \\
4116.47  &   V I  & 4.90E-01 & 0.275 & QC,S   ec                             \\
4116.56  &    V I & 1.47E-01 & 0.262 & QC,S & ec                                        \\
4131.99  &    V I &  8.51E-01  &  0.287 & QC,S & $-$30 \kmps{} emission. iP Cyg.          \\ 
4132.06  &    Fe{\small{I}} &  2.11E-01  &  1.608 & QC,S & the same                                            \\ 
4134.48  &    V I & 5.94E-01    & 0.301   & QC,S & ecs                                     \\
4158.62 ?&          &            &        & QC,S &   O II?                                            \\ 
4158.67 ?&          &            &        &  QC,S &  O II?                                           \\ 
4159.68  &    V I   &  1.86E-02  &  0.287 &  QC,S & ec redshifted.                         \\
4164.658 &    Nb I &  7.413E-01 &  0.049 & QC,S & in -30 \kmps{} emission line                    \\ 
4167.270 &    Gd I &   1.542E-01& 0.124  & QC,S &    P Cyg?                                        \\ 
4169.877 &    Ce II &  4.467E-01 &   0.536& QC,S &    P Cyg?                                       \\ 
4173.44  &          &            &         & QC,S &   iP Cyg                                           \\  
4178.85 &           &           &         &  QC,S &  forbidden Fe{\small{I}}?                                  \\ 
4181.93  &          &            &        &  QD,S &                                        \\
4184.07  &          &            &        &  QD,S &                                        \\
4185.15  &          &            &        &  QD,S &  S                                       \\
4191.09  &          &            &        &  QD,S &  S                                       \\ 
4198.30  &    Fe{\small{I}} &  1.91E-01  &  2.399 &   QC $>$ S & P Cyg                                  \\
4200.7   &          &            &        &   QC $>$ S & nebular?                            \\ 
4215.52  &    Sr II &  7.16E-01  &  0.0 & QC,S &                                              \\ 
4216.18  &    Fe{\small{I}} &  4.41E-04  &  0.0 &  QC,S & ec shifted blueward.                                 \\ 
4226.73  &    Ca I &  1.75E+00  &  0.0 &  QC,S & ecs shifted to the red \\
4227.43  &    26.00 & 1.84E+00  & 3.332 &  QC,S & P Cyg                                          \\
4233.15 &          &            &        &  QC,S &                                             \\ 
4237.27  &         &            &        &  QC,S &                                        \\
 4254.35  &   Cr I &   8.13E-01 & 0.000 &   QC,S &                                 \\
4259.31  &    V I &  6.76E-03  &  0.017 &   QC,S & P Cyg                    \\ 
4266.34  &         &           &        &   QC,S &                                    \\
4272.23 &          &            &        &  QC,S &                                             \\ 
4274.81  &    Cr I &  6.03E-01  &  0.0 &  QC,S & ec                               \\ 
4277.538 &    Ti{\small{I}}I &  1.816E-01 &  4.969 & QC,S &  P Cyg                                          \\ 
4289.73  &    Cr I &  4.27E-01  &  0.0 &  QC,S & ec                       \\ 
5889.95  &    Na I &  1.28E+00  &  0.0 &  QC,S & ecs       \\ 
5895.92  &    Na II &  6.40E-01  &  0.0 & QC,S &  ecs       \\ 
\end{longtable}
\end{onecolumn}
\end{appendix}
\end{document}